\documentclass[aps,prx,superscriptaddress,twocolumn,showpacs,noeprint]{revtex4-2}

\usepackage{graphicx,amsfonts,times,bm,amsmath,amssymb,verbatim,ifthen,braket,color,array,natbib,bm}
\usepackage[caption=false]{subfig}

\begin{document}

\title{Measurement-based time evolution for quantum simulation of fermionic systems}

\author{Woo-Ram Lee}
\affiliation{Department of Physics, Virginia Tech, Blacksburg, Virginia 24061, USA}

\author{Zhangjie Qin}
\affiliation{Department of Physics, Virginia Tech, Blacksburg, Virginia 24061, USA}

\author{Robert Raussendorf}
\affiliation{Department of Physics and Astronomy, University of British Columbia, Vancouver, BC V6T 1Z1, Canada}

\author{Eran Sela}
\affiliation{Department of Physics and Astronomy, Tel Aviv University, Tel Aviv 6997801, Israel}

\author{V.W. Scarola}
\email[Email address:]{scarola@vt.edu}
\affiliation{Department of Physics, Virginia Tech, Blacksburg, Virginia 24061, USA}

\begin{abstract}
Quantum simulation using time evolution in  phase estimation-based quantum algorithms can yield unbiased solutions of classically intractable models.  However, long runtimes open such algorithms to decoherence.  We show how measurement-based quantum simulation uses effective time evolution via measurement to allow runtime advantages over conventional circuit-based algorithms that use real-time evolution with quantum gates.  We construct a hybrid algorithm to find energy eigenvalues in fermionic models using only measurements on graph states.  We apply the algorithm to the Kitaev and Hubbard chains.  Resource estimates show a runtime advantage if measurements can be performed faster than gates, and graph states compactification is fully used. In this letter, we set the stage to allow advances in measurement precision to improve quantum simulation. 
\end{abstract}


\maketitle

\section{Introduction}

Unbiased quantum simulation \cite{Feynman1982,Georgescu2014} of intractable models aids in validating approximations.  Compelling open problems include the two-dimensional Hubbard model of the cuprates and, more generally, materials and quantum chemistry models \cite{Abrams1997,Abrams1999,Ortiz2001,Somma2002,AlanAspuru-Guzik2005,Kassal2010,Yung2012,Wecker2014,McClean2014,Wecker2015,Bauer2016,McClean2017,Babbush2018,Poulin2018,McArdle2020}.  Such interacting fermionic models are typically NP-hard because they suffer from the fermion sign problem \cite{Troyer2005} and are generally parameterized as: $H = \sum_{i,j} w_{ij} c_{i}^\dag c_{j}^{\vphantom{\dagger}} + \sum_{i,j,k,l} V_{ijkl} c_{i}^\dag c_{j}^\dag c_{k}^{\vphantom{\dagger}}c_{l}^{\vphantom{\dagger}}$, where $c_{j}^\dag$ creates a fermion in quantum state $j$ (a composite index for spin, lattice site, etc.) and $w_{ij} (V_{ijkl})$ is the single (two)-particle Hamiltonian matrix element. Since they are NP-hard, classical simulation time increases exponentially with particle number.  Unbiased quantum simulation of models captured by $H$ will therefore offer high impact benchmarks.  Variational quantum algorithms offer promise on near-term devices \cite{Moll2018} because they can be used to rigorously bound ground state energies. 

Recent work \cite{Ferguson2021} combines a variational quantum algorithm  with measurement-based quantum computing (MBQC) \cite{Raussendorf2001b,Raussendorf2003} for efficient management of variational ansatz states.  MBQC starts with a resource state, e.g., a graph state such as the square lattice cluster state (SLCS), formed by taking qubits aligned along the Pauli-$x$ direction and then entangling them pairwise with controlled-$Z$ gates. All quantum algorithms can then be executed using just single-qubit measurements on the resource state. MBQC-based variational algorithms \cite{Ferguson2021} can therefore use measurements to bound ground state energies.   

Phase-estimation-based quantum simulation algorithms \cite{Kitaev1995,Lloyd1996,Abrams1999,Berry2000,Higgins2007,Svore2014} can go beyond variational bounds to yield exact eigenfunctions and eigenvalues of $H$ for use in benchmarking excited state properties. In circuit-based quantum computing (CBQC), such algorithms take an input wave function $\vert \psi_{\text{I}} \rangle$ and repeatedly apply quantum gates to time-evolve $H$ with $M$ small time steps $\delta t_g$ to eventually extract information. Quantum algorithms based on this procedure yield an advantage over classical algorithms but for runtimes that increase exponentially with the required bit precision in, e.g., eigenvalues.  Long runtimes can be prohibitive \cite{Peruzzo2014,McClean2014,Wecker2015} if, for $N_g$ gates per time step, the qubits cannot be kept coherent for long execution times, $T_{\rm C}\sim M N_g \delta t_g$. 

We propose revisiting phase-estimation-based quantum simulation runtime from the MBQC perspective.  We consider the following regime: (i) A large number of qubits are available, (ii) the time taken for an accurate single-qubit measurement $\delta t_m$ can be made small enough to avoid decoherence of the resource state, and (iii) the entangling gates are performed in parallel mostly at the beginning. Assumption (iii) allows slow/error-prone entangling gates to be implemented and error corrected in a time that is negligible compared with the time to execute all measurements.  

\begin{figure}[t]
\begin{center}
\includegraphics[width=0.48\textwidth]{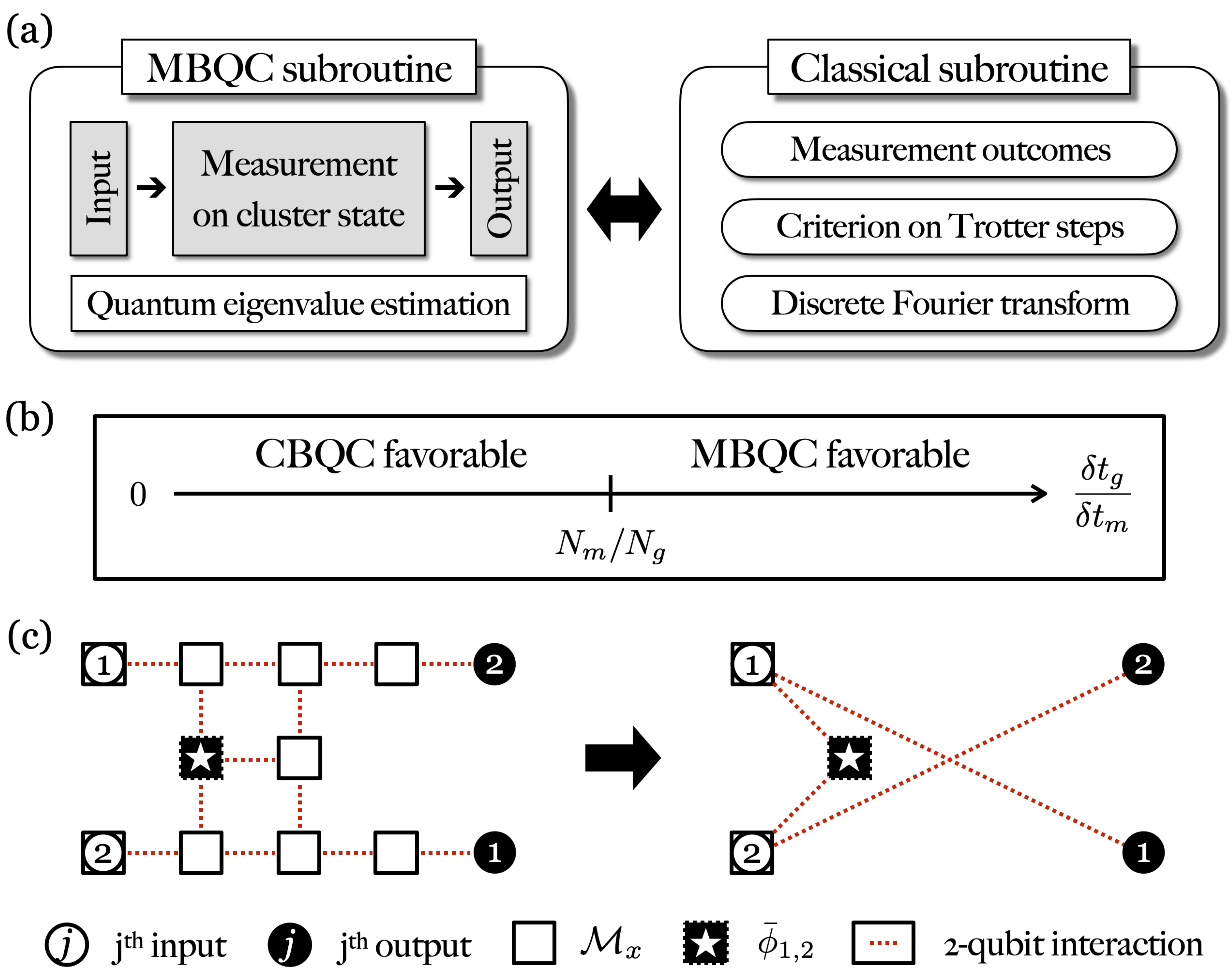}
\end{center}
\caption{
(a) Schematic for the hybrid quantum eigenvalue estimation algorithm. 
(b) Measurement-based quantum computing (MBQC)/circuit-based quantum computing (CBQC)-favorable regimes determined by the hardware-dependent parameter $\delta t_g/\delta t_m$. The point $N_m/N_g$ is obtained by setting the MBQC and CBQC runtimes to be the same $T_{\rm M}=T_{\rm C}$.
(c) Measurement-based effective time evolution used for a two-site Jordan-Wigner string where information flows from left to right.  Each box/circle hosts a single qubit entangled along red-dotted lines. Open (filled) circles are input (output) qubits. Open squares are Pauli-$x$ measurements that can be performed in parallel, and the dotted box around the central star indicates an adaptive measurement with an angle dictating the time step.  The left panel uses the square lattice cluster state (SLCS); 
the right panel is one of many   
equivalent graphs states that can be used instead, see Supplemental Material \cite{SM}.}
\label{fig_algorithm_hopping}
\end{figure}

In this letter, we explicitly map real-time evolution in CBQC (repeated application of gates that take a finite amount of time) to repeated measurement in MBQC~\cite{Raussendorf2003}.   To this end, we make the following advances: (i) We construct a route to use MBQC to effectively time-evolve $H$ using just single-qubit measurements.  We show that long effective time evolution corresponds to $M$ sequential measurement rounds in MBQC, thus requiring coherence among non-measured qubits for a total time, $T_{\rm M}\sim M N_m \delta t_m$, where $N_m$ is the number of measurements per round. (ii) We construct an example hybrid MBQC algorithm with a quantum phase-estimation-based  subroutine that yields exact eigenenergies: quantum eigenvalue estimation using an offline (classical) time series \cite{Somma2002,Somma2019}; see Fig.~\ref{fig_algorithm_hopping}(a). (iii) We apply the algorithm to solve the Kitaev \cite{Kitaev2000,Kitaev2009} and Hubbard \cite{Essler2005} chains because they can be solved exactly and can therefore be accurately checked as first implementations.  To compare $T_{\rm M}$ and $T_{\rm C}$ for our algorithm, we compute scaling of MBQC measurement time and precision costs as well as gate counts in an equivalent CBQC algorithm.

Our central finding is that MBQC can yield a runtime advantage over CBQC, i.e., $T_{\rm M}<T_{\rm C}$, by shifting the burden of requiring low $\delta t_g$ but high-fidelity gates in CBQC simulation to the requirement of low $\delta t_m$ and high single-qubit measurement precision in MBQC simulation. Figure~\ref{fig_algorithm_hopping}(b) summarizes our finding by showing that, if $\delta t_g/\delta t_m$ is large, MBQC will have shorter runtimes.  Here, $N_m/N_g$ is set by the algorithm.  We find that graph state compactification \cite{Hein2004} can yield hybrid MBQC algorithms with $N_m/N_g=1$. In this letter, therefore, we establishe a route to use improvements in quantum sensing \cite{Degen2017} to advance the state of the art in quantum simulation with effective time evolution.

\section{Measurement-Based Time Evolution}

Time evolution of Hamiltonians containing noncommuting terms $H_1$ and $H_2$ requires a decomposition.  The first-order Trotter-Suzuki decomposition is simplest \cite{Trotter1959,Suzuki1993}: $\exp[-i(H_1+H_2)t] = [\exp(-iH_1t/M) \exp(-iH_2t/M)]^M + \mathcal{O}[(t/M)^2]$. Here the time step $t/M$ is repeated $M$ times until the output state is converged within a tolerance $\delta_{\rm T}$~\cite{Comment1}, and $\hbar=1$.

To map between fermions and qubits in $H$, we use the Jordan-Wigner (JW) transformation~\cite{Jordan1928}: $c_j^\dag = \prod_{k=1}^{j-1} [-\sigma_z^{(k)}]$ $[\sigma_x^{(j)} + i \sigma_y^{(j)}]/2$, where $\sigma_a$ with $a\in\{x,y,z\}$ are the Pauli matrices. Long JW strings containing $N$ qubits can arise in certain models, e.g., those with long-range hopping/interaction in $H$.  Longer-range terms allow simulation of higher-dimensional fermionic models $H$ because they map to one-dimensional chains with long-range hopping and long-range interaction. Time evolution of a string requires the ability to execute nontrivial unitaries: $R_{a_1 a_2\cdots a_N}^{(1,2\cdots N)}(\theta) = \exp [-i(\theta/2)\prod_{j=1}^N\sigma_{a_j}^{(j)}]$, where $\theta$ is a rotation angle. 

The JW transformation enables construction of a time-to-angle mapping for MBQC simulation.  Figure~\ref{fig_algorithm_hopping}(c) shows an example measurement pattern needed for time evolution of a hop between neighboring sites, $c_1^\dag c_2^{\vphantom{\dag}}$ $+c_2^\dag c_1^{\vphantom{\dag}}$.  In the absence of the central measurement (star), the measurement pattern swaps information on qubits 1 and 2 \cite{Raussendorf2003,SM}. However, the additional adaptive measurement in the second round of measurements with $\bar{\phi}_{1,2}$ on the central qubit (star) incorporates results from the first round to yield \cite{Raussendorf2003}  $R_{zz}^{(1,2)}(\theta) \vert \psi_{\text{I}} \rangle$, where $\theta$ defines the part not relying on random measurement outcomes in $\bar{\phi}_{1,2}$.  This operation is a time propagator, and one can show, see Supplemental Material \cite{SM}, that, with a few more measurements, this measurement pattern effectively time-evolves a hop between sites 1 and 2.

Figure~\ref{fig_algorithm_hopping}(c) generalizes to time evolution of longer-range terms in $H$ on a larger SLCS using only $\mathcal{O}(1)$ adaptive measurements.  Consider, e.g., a long-range hop between sites $1$ and $N$: $c_1^\dag c_N^{\vphantom{\dag}}+c_N^\dag c_1^{\vphantom{\dag}}$.  To implement effective time evolution of the JW term, we must execute the unitary $R_{zz\cdots z}^{(1,2\cdots N)}(\theta)$ (and follow up with a few rotations on the end qubits).  This can be implemented with two rounds of measurements on $[(2N-1)^2$ $-(N-1)]$ qubits on the central area of the measurement pattern (excluding input and output qubits). The first round measures all but a central qubit, and the second round measures just the central qubit in an adaptive basis, see Supplemental Material \cite{SM}, thus showing a considerable simplification in implementing long JW strings.

The number of measurements and qubits needed for effective time evolution on a SLCS, e.g., the left side of Fig.~\ref{fig_algorithm_hopping}(c), can be significantly reduced. The Gottesman-Knill theorem~\cite{Gottesman1999} shows that all qubits with Pauli-$x$ measurements can be excluded since Clifford operations can be efficiently executed classically. After mathematically removing local Pauli measurements, the SLCS maps to a compactified cluster state (CCS)~\cite{Hein2004}. The mappings show that a much smaller graph state is needed. For example, the right side of Fig.~\ref{fig_algorithm_hopping}(c) shows an equivalent execution of $R_{zz}^{(1,2)}(\theta)$ (see Supplemental Material \cite{SM} for a proof), where the number of qubits (measurements) reduces from 12(10) to 5(3). In general, a CCS offers a reduction in measurement and qubit overhead for executing effective time evolution using $R_{z z\cdots z}^{(1,2\cdots N)}(\theta)$ by as much as $\mathcal{O}(N^2)$, depending on which CCS is chosen. We construct example time-evolution subroutines on SLCSs with the understanding that use of a CCS reduces the number of required qubits and measurements at the expense of modifying qubit connectivities which is efficiently programmable~\cite{Anders2006}.

\begin{figure*}[t]
\centering
\includegraphics[width=0.99\textwidth]{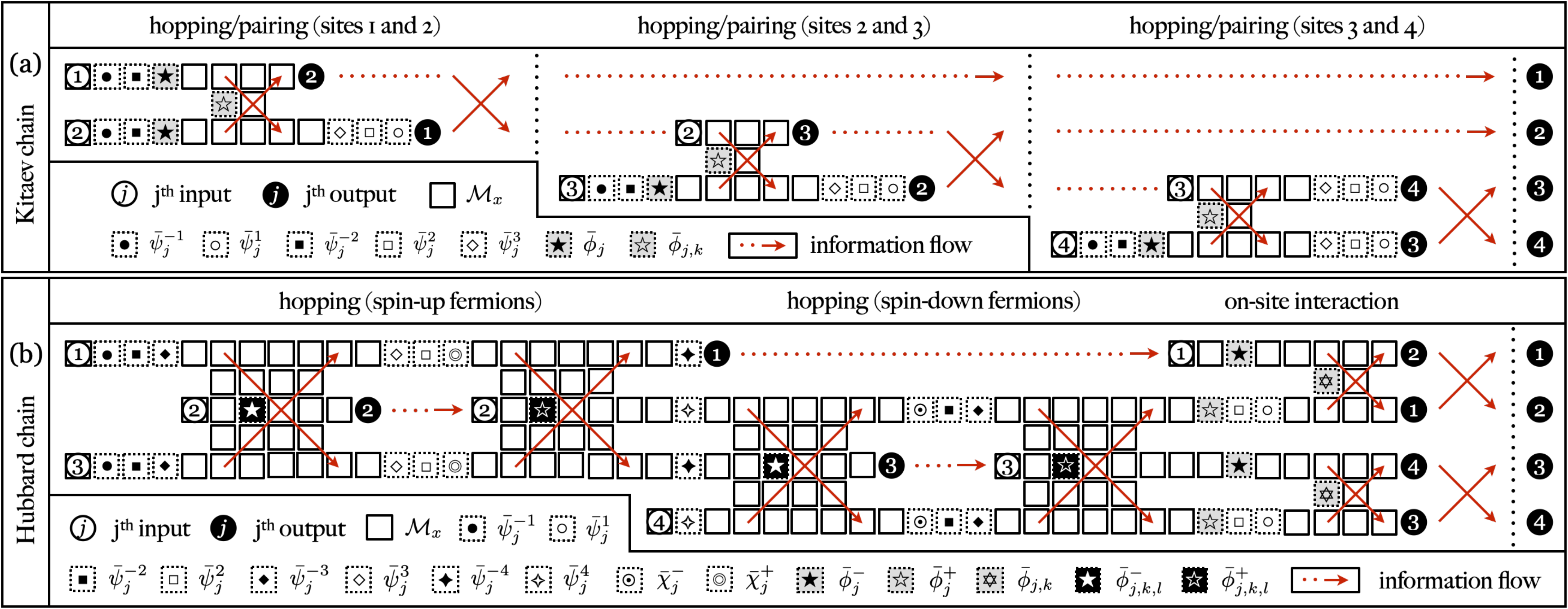}
\caption{
(a) Subroutine for implementing $\exp(-iH_{\text{K}}t) \vert \psi_{\text{I}} \rangle$ for four sites using only measurements on a square lattice cluster state (SLCS), as in Fig.~\ref{fig_algorithm_hopping}(c). Adaptive measurements are carried out with the angles defined in Eq.~\eqref{eq_Measurement_angle_Kitaev}. The indices $j,k\in\{1,2,3,4\}$ are assigned along the direction of information flow (red arrows).  Measurement angles denoted by stars execute effective time evolution, while other shapes denote measurements to perform rotations at the ends of the Jordan-Wigner (JW) strings. All qubits but inputs with Pauli-$x$ measurements (open boxes) can be removed in compactified cluster states (CCSs). (b) The same but for a two-site Hubbard chain with angles defined by Eq.~\eqref{eq_Measurement_angles_Hubbard} and $j,k,l\in\{1,2,3,4\}$.
}
\label{fig_measurement_pattern}
\end{figure*}

\section{Kitaev Chain}

We construct an MBQC subroutine for time evolution of an example model with noncommuting terms, the Kitaev chain \cite{Kitaev2000,Kitaev2009}: 
\begin{align}
H_{\text{K}}
& = w\sum_{j=1}^{N-1} ( - c_j^\dag c_{j+1}^{\vphantom{\dagger}} +  c_j^{\vphantom{\dagger}} c_{j+1}^{\vphantom{\dagger}} + {\rm H.c.}) - \mu \sum_{j=1}^N \delta n_j,
\label{eq_kitaev_chain_H}
\end{align}
where $w\geq 0$ is the hopping and pairing energy, $\mu\geq 0$ is the chemical potential, and $\delta n_j = c_j^\dag c_j^{\vphantom{\dagger}} - 1/2$. The ground state exhibits a quantum phase transition at $\mu=2w$ between a non-topological strong-coupling phase ($\mu > 2w$) and a topological weak-coupling phase ($\mu < 2w$).

We map fermions to qubits to construct both circuit- and measurement-based time propagators. The JW transformation maps $H_{\text{K}}$ to the quantum Ising model. The first-order Trotterized form of $\exp(-iH_{\text{K}}t)$ is
\begin{align}
\left[ \prod_{j,k} R_{xx}^{(j,j+1)}(-2\phi_M) R_z^{(k)}(-2g_{\mu}\phi_M) \right]^M,
\label{Quantum_gate_Kitaev}
\end{align}
where $g_{\mu}=\mu/(2w)$ and
\begin{align}
\phi_M =\frac{wt}{M}
\label{eq_time_angle_mapping}
\end{align}
is a measurement angle. Equation~\eqref{Quantum_gate_Kitaev} can be implemented in two different ways: using real-time evolution in CBQC or effective time evolution in MBQC, where $M$ dictates the circuit or measurement depth, respectively. Equation~\eqref{eq_time_angle_mapping} is central because it maps real time $t$ to measurement angle.

We use the stabilizer formalism to map Eq.~\eqref{Quantum_gate_Kitaev} to effective time evolution in MBQC.  Figure~\ref{fig_measurement_pattern}(a) shows the measurement pattern implementing Eq.~\eqref{Quantum_gate_Kitaev} to time-evolve input qubits 1-4 (open circles) with just single-qubit measurements. The measurement angles in the $x$-$y$ plane are
\begin{align}
\bar{\phi}_{j,k} & = 2P_{\bar{\phi}_{j,k}} \phi_M,
~~~~
\bar{\phi}_{j} = -P_{\bar{\phi}_j} (2g_{\mu}\phi_M + \gamma),
\nonumber\\
\bar{\psi}_{j}^{r} & =  P_{\bar{\psi}_{j}^{r}} \psi^{r},
\label{eq_Measurement_angle_Kitaev}
\end{align}
where $\psi^r\in\{\pm\alpha,\pm\beta,\gamma\}$ for $r=\pm1,\pm2,3$, $-\alpha = \beta = \gamma =$ $\pi/2$, and $P_\theta = (-1)^{S_\theta^{\text{K}}}$. Here, $S_\theta^{\text{K}}$ accumulates all measurement outcomes during single-qubit measurements and is derived in the Supplemental Material \cite{SM}. The measurement outcomes are also used for offline processing with a byproduct operator, see Supplemental Material \cite{SM}, that defines the basis for interpreting output measurements.

The left, middle, and right panels depict measurements [stars in Fig.~\ref{fig_measurement_pattern}(a)] that entangle input qubits 1-2, 2-3, and 3-4, respectively. The measurement pattern in Fig.~\ref{fig_measurement_pattern}(a) and Eq.~\eqref{eq_Measurement_angle_Kitaev} define the full effective time-evolution algorithm for a Kitaev chain of any $N$ or $M$ because additional panels in Fig.~\ref{fig_measurement_pattern}(a) can be concatenated, see Supplemental Material \cite{SM}. The red dots and arrows show information flow for use in concatenation.

\section{Hubbard Chain}

We now turn to the Hubbard chain \cite{Essler2005} with a longer JW string and an important interaction term:
\begin{align}
H_{\text{H}}
= - w \sum_{j=1,\sigma}^{N-1} (c_{j,\sigma}^\dag c_{j+1,\sigma}
^{\vphantom{\dagger}}+ {\rm H.c.}) + U \sum_{j=1}^N n_{j,\uparrow} n_{j,\downarrow},
\label{eq_Hubbard_chain}
\end{align}
where $\sigma\in\{\uparrow,\downarrow\}$, $U$ is the Hubbard interaction strength, and $n_{j,\sigma} = c_{j,\sigma}^\dag c_{j,\sigma}^{\vphantom{\dagger}}$. To map fermions to qubits, we introduce \cite{Camp1974} spinless fermion operators: $\tilde{c}_{2j-1} = c_{j,\uparrow}$ and  $\tilde{c}_{2j} = c_{j,\downarrow}$. The JW mapping then leads to an equivalent qubit Hamiltonian: $(w/2) \sum_{j=1}^{2N-2}(\sigma_x^{(j)}$ $\sigma_z^{(j+1)} \sigma_x^{(j+2)} + \sigma_y^{(j)} \sigma_z^{(j+1)} \sigma_y^{(j+2)})$ $+ (U/4) \sum_{j=1}^{N} (\mathbb{I}_{2j-1} + \sigma_z^{(2j-1)}) (\mathbb{I}_{2j} + \sigma_z^{(2j)})$, where the JW strings used for the hopping terms need a three-qubit entangling gate, and $\mathbb{I}={\rm diag}(1,1)$. The first-order Trotterized form of $\exp(-iH_{\text{H}}t)$ is
\begin{align}
& \Bigg[ \prod_{j,k} R_{zz}^{(2j-1,2j)}(g_U\phi_M) R_z^{(2j-1)}(g_U\phi_M) R_z^{(2j)}(g_U\phi_M) 
\nonumber\\
& ~~~~\times R_{xzx}^{(k,k+1,k+2)}(\phi_M) R_{yzy}^{(k,k+1,k+2)}(\phi_M) \Bigg]^M,
\label{Quantum_gate_Hubbard}
\end{align}
where $g_U= U/(2w)$.

Equation~\eqref{Quantum_gate_Hubbard} can be used in CBQC or mapped to single-qubit measurements in MBQC. Figure~\ref{fig_measurement_pattern}(b) depicts the $N=2$ measurement pattern for Eq.~\eqref{Quantum_gate_Hubbard} with measurement angles:
\begin{align}
& \bar{\phi}_{j,k,l}^{\pm}
= -P_{\bar{\phi}_{j,k,l}^{\pm}}\phi_M,
~~
\bar{\phi}_{j,k} = -P_{\bar{\phi}_{j,k}}g_U\phi_M,
~~
\bar{\psi}_{j}^{r} 
= P_{\bar{\psi}_{j}^{r}}\psi^r,
\nonumber\\
& \bar{\phi}_{j}^{\pm} 
= \pm P_{\bar{\phi}_{j}^{\pm}}[g_U\phi_M + (1\pm 1)\gamma/2],
~~
\bar{\chi}_{j}^{\pm} 
= \pm P_{\bar{\chi}_{j}^{\pm}}(\lambda+\alpha),
\label{eq_Measurement_angles_Hubbard}
\end{align}
where $\psi^r\in\{\pm\alpha,\pm\beta,$ $\pm\gamma,\pm\lambda\}$ for $r=\pm1,\pm2,\pm3,\pm4$, $\lambda = \alpha$, and $P_\theta = (-1)^{S_\theta^{\text{H}}}$.  $S_\theta^{\text{H}}$ is derived in Ref.~\cite{SM}.  The Hubbard chain measurement pattern can also be concatenated to time-evolve larger $N$ or $M$~\cite{SM}, and overhead can be significantly reduced in a CCS.  

Both examples, Eqs.~\eqref{eq_Measurement_angle_Kitaev} for $H_{\text{K}}$ and Eqs.~\eqref{eq_Measurement_angles_Hubbard} for $H_{\text{H}}$, demonstrate constraints on effective time evolution.  Long effective time evolution from a larger number of Trotter steps in MBQC corresponds to smaller measurement angles since $\phi_M \propto 1/M$.  Repeated small-angle measurements (long effective time evolution) in MBQC therefore require improvements in qubit measurement precision as opposed to faster gates in CBQC.

\begin{figure}[t]
\begin{center}
\includegraphics[width=0.44\textwidth]{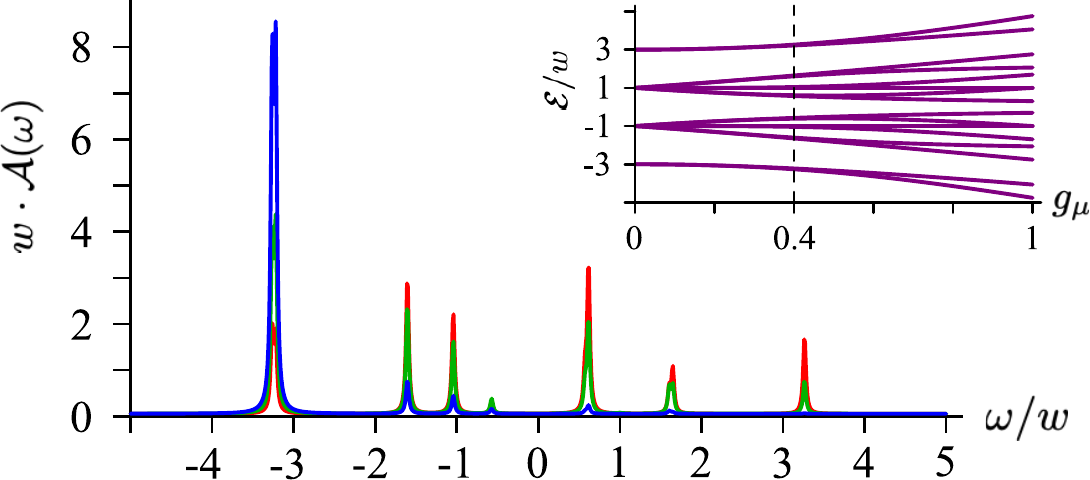}
\end{center}
\caption{Main: Simulation using Eq.~\eqref{eq_Spectral_function}, where peaks reveal the exact eigenenergies of the four-site Kitaev chain with $g_{\mu}=0.4$, $\eta/w = 0.02$, and $\delta\omega/w = 0.01$.  Trotter error is $\delta_{\rm T} = 10^{-2}$ for $M<8500$, and $L=1272$ is chosen for clarity. $\vert \psi_{\text{I}}\rangle$ is chosen to be the ground state at $g_{\mu}=0$. The blue line indicates the error-free case, and the green and red lines plot the impact of random perturbations [45-56$\%$ (green); 70-82$\%$ (red)] in the measurement angles $\bar{\phi}_j$ and $\bar{\psi}_j^{3}$. Inset: Eigenenergies of Eq.~\eqref{eq_kitaev_chain_H}, where the energies touching the dashed line match the peak positions in the main panel.}
\label{fig_spectrum}
\end{figure}

\section{Eigenvalue Estimation}

To demonstrate resource requirements, we construct a minimal hybrid quantum eigenvalue estimation algorithm by combining MBQC subroutines with an offline time-series estimator [Fig.~\ref{fig_algorithm_hopping}(a)]. A $|\psi_{\text{I}}\rangle$ close to a desired eigenstate is fed into the MBQC time-evolving subroutine yielding $\langle \psi_{\text{I}} | e^{-iHt} |\psi_{\text{I}}\rangle$ if the output qubits are measured using quantum state tomography (or an ancilla qubit \cite{Somma2002}) to find the wavefunction phase relative to the input qubit basis. The MBQC output is obtained $L$ times and used in a classical discrete Fourier transform:
\begin{align}
\mathcal{A}(\omega_m)
& = \frac{\delta t}{\pi} \sum_{n=0}^{L-1} {\rm Re} \big[ e^{(i\omega_m-\eta) t_n} \langle \psi_{\text{I}} | e^{-iHt_n} |\psi_{\text{I}}\rangle \big],
\label{eq_Spectral_function}
\end{align}
where we define $t_n = n\delta t$, $\omega_m = m\delta\omega$ ($n,m=0,1,\cdots,L-1$) in units of $\delta t$ and $\delta\omega$ satisfying $\delta\omega \delta t = 2\pi/L$.  Peaks in $\mathcal{A}(\omega)$ yield eigenvalues of $H$ to within $\delta_{\rm T}$. We introduce the broadening parameter $\eta > 0$ for visualization of Lorentzian peaks and as a proxy for experimentally limited resolution.  

The main panel in Fig.~\ref{fig_spectrum} shows a demonstration result from a simulation using $H_{\text{K}}$ in Eq.~\eqref{eq_Spectral_function}, where several eigenvalues are returned as peaks. One can show, see Supplemental Material \cite{SM}, that peak centers are intact while peak weights are shifted for certain types of measurement errors. Figure~\ref{fig_spectrum} uses large $L$ and $M$ for clarity but in practice, $L$ and $M$ can be lowered.  They are minimized by restricting the search to just the ground state energy, while three independent algorithm input parameters $\delta \omega$, $L$, and $M$ must be chosen to meet three tolerances: (i) $\delta \omega$ should be smaller than  $\eta$ to resolve peak structure, (ii) A sum rule tolerance  $\delta_{\rm F} > \big|1 - \delta\omega \sum_{m=0}^{L-1} \mathcal{A}(\omega_m)\big|$ sets $L$, and (iii) $M$ is set by requiring $\delta_{\rm T}$ to be much smaller than the first spectral gap.

\section{Measurement Precision}

The number of Trotter steps yields the measurement depth and sets $\phi_M$.  Large $M$ improves Trotter accuracy at the expense of requiring improved measurement precision (small $\phi_M$).  To estimate the minimum $M$ needed to obtain ground state energies, we consider $H_{\text{K}}$ with $g_{\mu}=0.01-0.4$.  We find empirically, see Supplemental Material \cite{SM}, that, for each $n$ in Eq.~\eqref{eq_Spectral_function}, the minimum $M$ varies from $1.8\times10^3$ ($g_{\mu}=0.01$) to $7.8\times10^4$ ($g_{\mu}=0.4$) to resolve the ground state energy of $H_{\text{K}}$ to within $1\%$ of the spectral gap ($\delta_{\rm T} = 10^{-2}$) for $\eta=0.02 w$, $\delta\omega=0.01w$, $L=46$, and $N=4$. We have checked $N\leq8$ with other $\eta$, $\delta\omega$, and $L$ combinations and obtained similar results for $M$.  In general, the $M$ needed will depend on the model, model parameters, tolerances, and scales as $\mathcal{O}((N t_n)^2 \delta_{\rm T}^{-1})$ \cite{Childs2021}, thus implying that the required measurement depth and precision to execute effective MBQC time evolution can become demanding~\cite{Comment2}.

Given bounds on $M$, we can estimate measurement precision requirements for $H_{\text{K}}$. Here, $\phi_M$ depends on $n$.  The largest measurement angle (in units of $2\pi$) needed to implement Eq.~\eqref{eq_Spectral_function} with Eq.~\eqref{eq_Measurement_angle_Kitaev} is $\chi_{L-1}$, where $\chi_n\equiv nw/(\delta\omega L M)$.  We empirically find, see Supplemental Material \cite{SM} (far from the critical point at $g_{\mu}=1$), $\chi_n\lesssim 0.14$, thus allowing the use of Eq.~\eqref{eq_time_angle_mapping}.  The smallest measurement angle increment needed in Eq.~\eqref{eq_Measurement_angle_Kitaev} is $g_{\mu}\chi_n$.  We find $g_{\mu}\chi_n\gtrsim4.8\times10^{-4}$ for all $g_\mu<1$ and $n$. We therefore see that a large $M$ requires small angle measurements as we implement effective time evolution. 

\begin{table}[t]
\caption{
\label{table_resources}
Resources for a single time step in Eq.~\eqref{eq_Spectral_function} computed by counting and concatenation, see Supplemental Material \cite{SM}, in three scenarios(rows): (i) MBQC on an SLCS including all Pauli-$x$ and adaptive measurements, (ii) MBQC on a CCS with the least number of measurements, and (iii) CBQC.  In (i) and (ii), measurements on input/output qubits are not counted.  (ii) and (iii) show the same scaling ($N_m/N_g=1$) for two different experimental processes, measurements and two-qubit gates.}
\begin{ruledtabular}
\begin{tabular}{lll}
Approach  & $H_{\text{K}}$ & $H_{\text{H}}$ \\ [0.5ex]
\hline
SLCS Measurements, $N_m$ & $ (17 N - 10)M$ & $(156 N - 144)M$ \\
\hline
CCS Measurements, $N_m$ & $(7N-1)M$ & $(34N-32)M$  \\
\hline
Circuit-based Gates, $N_g$ & $(7N -1)M$ & $(34N - 32)M$ \\
\end{tabular}
\end{ruledtabular}
\end{table}

\section{Measurement and Qubit Overhead}

The measurement subroutines defined by Eqs.~\eqref{eq_Measurement_angle_Kitaev} and \eqref{eq_Measurement_angles_Hubbard} allow estimates of resource requirements in our hybrid quantum eigenvalue estimation algorithm. Table~\ref{table_resources} shows, see Supplemental Material \cite{SM}, that, for the local models considered here, a CCS will have $N_m/N_g=1$. However, with nonlocal qubit terms, e.g., for nonlocal hopping in $H$, MBQC with a CCS will have an $\mathcal{O}(N)$ advantage in measurement vs gate counts in CBQC unless nonlocal gates are used to implement the JW strings \cite{Hastings2014c}.  The number of qubits needed is $\mathcal{O}(M)$ larger for MBQC than for CBQC.  MBQC qubit overhead can be lowered by re-entangling measured qubits \cite{Raussendorf2001b}.

\section{Discussion}

Our demonstration algorithms show that unbiased quantum simulation using effective time evolution is possible using only single-qubit measurements on graph states.  We find that long MBQC effective time evolution for use in quantum simulation requires high measurement precision to be useful in benchmarking approximate classical algorithms. Alternative time-evolution decompositions \cite{Berry2014,Berry2016,Poulin2018,Childs2018b,Low2017} will lower overhead.  

MBQC offers advantages in systems with slow/error-prone entangling gates \cite{Briegel2009a}, e.g., photonics \cite{Kok2007,Pick2021} and atoms in optical lattices \cite{Raussendorf2001b}.  In the latter case, parallelized collisional gates encoded large SLCSs in long-lived atomic hyperfine states \cite{Mandel2003}.  Recent progress in single-site measurements \cite{Wang2015} and control \cite{Wang2016} allow optical lattice implementation of MBQC effective time-evolution algorithms.  

The above algorithms have a low error threshold \cite{Aliferis2006,Dawson2006a}. An improvement with higher thresholds is available \cite{Raussendorf2006,Raussendorf2007b}.  The above algorithms can also be used in conjunction with an adaptive Bayesian algorithm (instead of a time series) in eigenvalue estimation learning certain types of error \cite{Wiebe2016,OBrien2019}.  

Finally, applications to higher-dimensional fermionic models are highly desired. Nearest neighbor hoppings/interactions in a higher-dimensional fermionic lattice can be mapped to long-range hoppings/interactions in a chain~\cite{Somma2002}. After mapping, our hybrid MBQC algorithm can be applied to the chain at the expense of increasing the length of JW strings.

\begin{acknowledgments}
We acknowledge support from ARO (W911NF2010013). W.-R.L., Z.Q., and V.W.S. acknowledge support from AFOSR (FA9550-18-1-0505, FA9550-19-1-0272).
\end{acknowledgments}

\bibliography{references} 

\begin{thebibliography}{61}
\bibitem [{Com({\natexlab{a}})}]{Comment1}%
  \BibitemOpen
  \bibinfo {note} {Specifically, we have $||\exp[-i(H_1+H_2)t] -
  [\exp(-iH_1t/M)$ $\exp(-iH_2t/M)]^M|| < \delta_{\rm T}$, where $||A||$ is the
  spectral norm, i.e., the largest singular value of $A$.}\BibitemShut {Stop}%
\bibitem [{Com({\natexlab{b}})}]{Comment2}%
  \BibitemOpen
  \bibinfo {note} {$\delta_{\rm T}$ is a strict upper bound on Trotter error,
  and estimates of energies with lower $M$ are possible. Furthermore, $M$ will
  decrease substantially for higher-order Trotter-Suzuki decompositions: $M
  \sim \mathcal{O}((N t_n)^{1+1/p}\delta_{\rm T}^{-1/p})$ for the order of
  $p$.}\BibitemShut {Stop}%
\end{thebibliography}

\end{document}


\title{Supplemental Material:\\
Measurement-Based Time Evolution for Quantum Simulation of Fermionic Systems}

\author{Woo-Ram Lee}
\affiliation{Department of Physics, Virginia Tech, Blacksburg, Virginia 24061, USA}

\author{Zhangjie Qin}
\affiliation{Department of Physics, Virginia Tech, Blacksburg, Virginia 24061, USA}

\author{Robert Raussendorf}
\affiliation{Department of Physics and Astronomy, University of British Columbia, Vancouver, BC V6T 1Z1, Canada}

\author{Eran Sela}
\affiliation{Department of Physics and Astronomy, Tel Aviv University, Tel Aviv 6997801, Israel}

\author{V.W. Scarola}
\email[Email address:]{scarola@vt.edu}
\affiliation{Department of Physics, Virginia Tech, Blacksburg, Virginia 24061, USA}

\maketitle

\section{Measurement pattern for effective time evolution of Jordan-Wigner string}
\label{sec_measurement_pattern_JW_string}

In this section, we prove that, as stated in the main text, the measurement pattern in the left side of Figure~1(b) effectively time-evolves a Pauli-$z$ string between sites 1 and 2.  Pauli-$z$ strings are needed for Jordan-Wigner (JW) strings. We also prove that Figure~1(b) generalizes to time evolution of longer-range strings on a larger square lattice cluster (SLCS) state using only $\mathcal{O}(1)$ adaptive measurements. 

We start with the definition of a SLCS.  Consider a connected subset of a square lattice $\mathcal{L}_2$ with vertices $\mathcal{V}(\mathcal{L}_2)$ and edges $\mathcal{E}(\mathcal{L}_2)$.  The SLCS is defined by \cite{Briegel01}:
\begin{align}
|G_{\rm SLCS}\rangle = \prod_{(j,k)\in \mathcal{E}(\mathcal{L}_2)} U_{\rm CZ}^{(j,k)} \prod_{l\in \mathcal{V}(\mathcal{L}_2)} |+\rangle_{x,l},
\label{cluster_state}
\end{align}
where $|\pm\rangle_{x,j} = (|0\rangle_j \pm |1\rangle_j)/\sqrt{2}$ are the Pauli-$x$ eigenstates, entangled by the controlled-$Z$ gate $U_{\rm CZ}^{(j,k)}$. The SLCS satisfies the eigenvalue equation $K_j |G_{\rm SLCS}\rangle = |G_{\rm SLCS}\rangle$ with the stabilizer given by
\begin{align}
K_j = X_j \prod_{k\in N_j} Z_k.
\end{align}
Here the Pauli matrices are abbreviated by $(X, Y, Z) \equiv (\sigma_x, \sigma_y, \sigma_z)$, and $N_j$ indicates the nearest neighbors of site $j$.

An SLCS can serve as a platform to implement unitary gates equivalent to circuit-based quantum computing (CBQC) gates.  The SLCS is composed of three sections: an input, a body, and an output. The number of input qubits is the same as the number of qubits defining the qubit Hamiltonian.  The number of output qubits is the same as the number of input qubits.  A central idea in measurement-based quantum computing (MBQC) is that a sequence of single-qubit measurements on the input and body sections trigger information flow from the input to the output~\cite{Raussendorf01}. The measurement pattern in the body determines which gate operation is encoded. Qubits in the input and body sections may be measured in an adaptive/non-adaptive basis in the $x$-$y$ plane. Here the measurement basis is written by $\mathcal{B}_j(\bar{\phi}_j) = \{ |\psi_j^{\hat{r}_j}\rangle_{s_j=0}, |\psi_j^{\hat{r}_j}\rangle_{s_j=1} \}$, where we define the single-qubit wavefunction  $|\psi_j^{\hat{r}_j}\rangle_{s_j} = [ |0\rangle_j + (-1)^{s_j} e^{i\bar{\phi}_j} |1\rangle_j ] / \sqrt{2}$ with the measurement angle $\bar{\phi}_j$ (along with the vector $\hat{r}_j = \hat{x} \cos\varphi_j + \hat{y} \sin\varphi_j$) and the random measurement outcomes $s_j\in\{0,1\}$.

After measurements we must define the measurement basis using a byproduct operator that returns equivalent CBQC gates.  We consider the projected SLCS: $|G_{\rm proj}\rangle = \prod_j\mathcal{P}_{s_j}^{(j)}(\bar{\phi}_j) |G_{\rm SLCS}\rangle$, where $j$ spans all sites in the input and body sections, and the projector is defined by $\mathcal{P}_{s_j}^{(j)}(\bar{\phi}_j) = [\mathbb{I}_j + (-1)^{s_j} \hat{r}_j \cdot \vec{X}_j]/2$. Here, $\mathbb{I} = {\rm diag}(1,1)$, and $\vec{X} = X\hat{x} + Y\hat{y} + Z\hat{z}$. In Ref.~\cite{Raussendorf03}, it was shown that $|G_{\rm proj}\rangle$ is governed by a set of eigenvalue equations:
\begin{align}
X_j \otimes (U_g X_k U_g^\dag) |G_{\rm proj}\rangle
& = (-1)^{\lambda_{x,j}} |G_{\rm proj}\rangle,
\label{Eigenvalue_equation_x}
\\
Z_j \otimes (U_g Z_k U_g^\dag) |G_{\rm proj}\rangle
& = (-1)^{\lambda_{z,j}} |G_{\rm proj}\rangle,
\label{Eigenvalue_equation_z}
\end{align}
where $j$ ($k$) is an index for the input (output) qubits, spanning from $1$ to $N$. Once we find $U_g$, $\lambda_{x,j}$, $\lambda_{y,j}$ in Eqs.~\eqref{Eigenvalue_equation_x} and \eqref{Eigenvalue_equation_z}, the output wavefunction has the connection to the input: $|\psi_{\rm O}\rangle = U_g U_\Sigma |\psi_{\rm I}\rangle$ up to the $U(1)$ phase factor.  The byproduct operator is defined by
\begin{align}
U_\Sigma = \prod_{j=1}^N Z_j^{s_j+\lambda_{x,j}} X_j^{\lambda_{z,j}},
\end{align}
which adjusts the output qubit basis. $U_g$ and $U_\Sigma$ can be switched using Pauli propagation relations, e.g., $X Z = - Z X$ and $Y Z = - Z Y$. Then, the output wavefunction can be refined by $|\psi'_{\rm O}\rangle = U_\Sigma |\psi_{\rm O}\rangle = U_g |\psi_{\rm I}\rangle$ to retain the right basis. If we match $U_g$ as in CBQC, we can derive the connection between the measurement angles and the model parameters.

We now apply the above formalism to the construction of JW strings.  As mentioned in the main text, time evolution of an $N$-site JW string requires the ability to execute $N$-qubit rotation gates: 
\begin{align}
R_{a_1 a_2\cdots a_N}^{(1,2\cdots N)}(\theta) = \exp \bigg(-i\frac{\theta}{2}\prod_{j=1}^N\sigma_{a_j}^{(j)}\bigg),
\label{N_qubit_rotation}
\end{align}
where $\theta$ is a rotation angle and $a_j\in\{x,y,z\}$. It turns out that Eq.~\eqref{N_qubit_rotation} has the decomposition:
\begin{align}
\Bigg(\prod_{j=1}^N U_{\rm rot}^{(j)\dag}\Bigg) R_{zz\cdots z}^{(1,2\cdots N)}(\theta) \Bigg(\prod_{k=1}^N U_{\rm rot}^{(k)}\Bigg),
\label{N_qubit_rotation_decomp}
\end{align}
where $U_{\rm rot}^{(j)} = R_x^{(j)}(\gamma_j)R_z^{(j)}(\beta_j)R_x^{(j)}(\alpha_j)$ executes the Euler rotation of a qubit at site $j$ to adjust the qubit basis. In the following two subsections, we use the stabilizer formalism to derive the mathematical details of the measurement patterns for $R_{zz}^{(1,2)}(\theta)$ and $R_{zzz}^{(1,2,3)}(\theta)$ which play a central role in MBQC simulation of the Kitaev and Hubbard chains (A proof for $N=2$ is outlined in Ref.~\cite{Raussendorf03}, but we include this case here for consistency).  

The full JW string takes $R_{zz\cdots z}^{(1,2\cdots N)}(\theta)$ and then 
applies single-qubit rotations to the end qubits (1 and $N$). These rotations can also be implemented with measurements.  To implement single-qubit rotation gates with Euler angles $(\alpha,\beta,\gamma)$, we use \cite{Raussendorf03}: $U_g = U_{\rm rot}$, $U_\Sigma = Z^{s_1+s_3} X^{s_2+s_4}$, with the measurement angles given by 
\begin{align}
\bar{\psi}^1 = -(-1)^{s_1}\alpha,
~~
\bar{\psi}^2 = -(-1)^{s_2}\beta,
~~
\bar{\psi}^3 = -(-1)^{s_1+s_3}\gamma,    
\end{align}
where the index $j$ in $\bar{\psi}^j$ and $s_j$ is defined along the 5-qubit cluster chain with the input (output) qubit at $j=1(5)$.  In the following two sections, we construct the measurement patterns needed to implement $R_{zz}^{(1,2)}(\theta)$ and $R_{zzz}^{(1,2,3)}(\theta)$ with the understanding that we must follow up with single-qubit rotations to the end qubits as shown in  Sec.~\ref{sub_sec_subroutines_fermions}.

\begin{figure}[b]
\begin{center}
\includegraphics[width=0.83\textwidth]{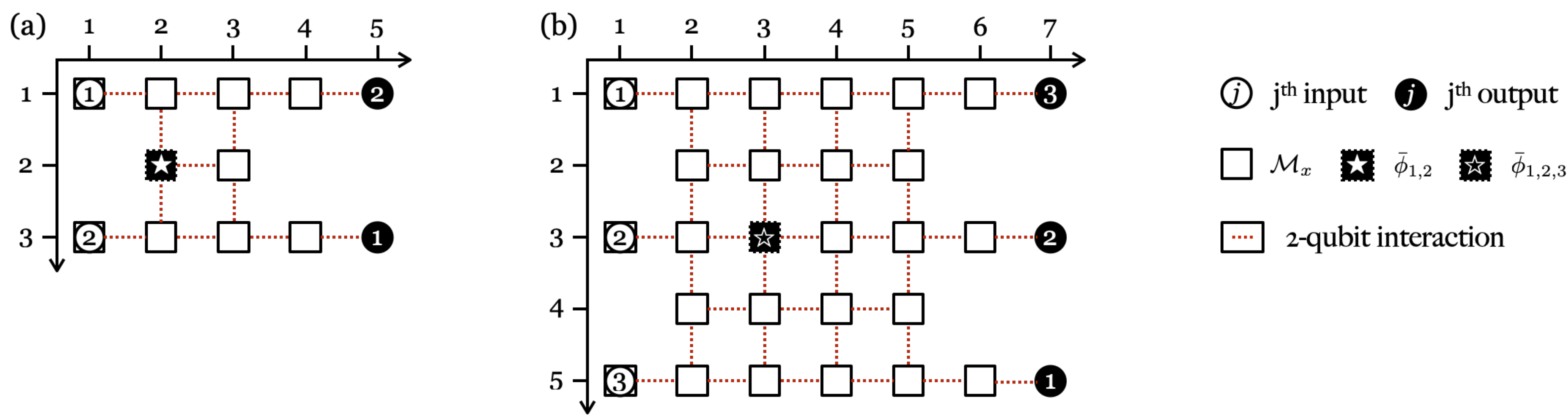}
\end{center}
\caption{Coordinate systems for assigning qubits in the measurement patterns for (a) 2-qubit rotation gate, (b) 3-qubit rotation gate.}
\label{fig_coordinate_systems}
\end{figure}

\subsection{Two-site ($N=2$) Jordan-Wigner string}

In this section, we prove that the measurement pattern in the left side of Figure~1(b) effectively time-evolves a Pauli-$z$ string between sites 1 and 2 that is needed in a JW string.  We build a SLCS with 12 qubits [see Figure~\ref{fig_coordinate_systems}(a)], defined by the composite stabilizers: 
\begin{align}
K_{(1,n)} K_{(2,2)} K_{(3,\bar{n})} K_{(5,\bar{n})},
~~
K_{(2,n)} K_{(3,2)} K_{(4,\bar{n})},
~~
K_{(3,2)} K_{(4,1)} K_{(4,3)},
\nonumber
\end{align}
where $n=1$ if $\bar{n}=3$ and vice versa. Carrying out the first round of Pauli-$x$ measurements on qubits at site (2,1), (3,1), (4,1), (3,2), (2,3), (3,3), (4,3), we find that the first-round projected SLCS $|G_{\rm proj}\rangle$ satisfies:
\begin{align}
Z_{(2,2)} Z_{(5,1)} Z_{(5,3)} |G_{\rm proj}\rangle & = (-1)^{s_{(3,2)}+s_{(4,1)}+s_{(4,3)}} |G_{\rm proj}\rangle,
\label{EigenvalueEq_N2_ZZZ}
\\
X_{(1,n)} X_{(2,2)} X_{(5,\bar{n})} |G_{\rm proj}\rangle & = (-1)^{s_{(3,\bar{n})}} |G_{\rm proj}\rangle,
\label{EigenvalueEq_N2_XXX}
\\
Z_{(1,n)} Z_{(5,\bar{n})} |G_{\rm proj}\rangle & = (-1)^{s_{(2,n)}+s_{(3,2)}+s_{(4,\bar{n})}} |G_{\rm proj}\rangle.
\label{EigenvalueEq_N2_ZZ}
\end{align}
Combining Eqs.~\eqref{EigenvalueEq_N2_ZZZ} and \eqref{EigenvalueEq_N2_XXX} leads to:
\begin{align}
X_{(1,n)} R_z^{(2,2)}(\bar{\phi}_{1,2}) X_{(2,2)} R_z^{(2,2)}(-\bar{\phi}_{1,2}) R_{zz}^{((5,1),(5,3))}(-P_{\bar{\phi}}\bar{\phi}_{1,2}) X_{(5,\bar{n})} R_{zz}^{((5,1),(5,3))}(P_{\bar{\phi}}\bar{\phi}_{1,2}) |G_{\rm proj}\rangle 
= Q_{\bar{\phi}} |G_{\rm proj}\rangle,
\label{EigenvalueEq_N2_XRXRRXR}
\end{align}
for arbitrary angle $\bar{\phi}_{1,2}$. Here we define $P_{\bar{\phi}} = (-1)^{s_{(3,2)}+s_{(4,1)}+s_{(4,3)}}$ and $Q_{\bar{\phi}} = (-1)^{s_{(3,\bar{n})}}$. It turns out that Eq.~\eqref{EigenvalueEq_N2_ZZ} can be recast to become a counterpart of Eq.~\eqref{EigenvalueEq_N2_XRXRRXR}. Carrying out the second round of measurement on qubit at site (2,2) in the basis $\mathcal{B}_{(2,2)}(\bar{\phi}_{1,2})$, we find that the second-round projected SLCS $|G'_{\rm proj}\rangle$ satisfies: 
\begin{align}
\left\{
\begin{array}{c}
X_{(1,n)}
\\
Z_{(1,n)}
\end{array}
\right\}
R_{zz}^{((5,1),(5,3))}(-P_{\bar{\phi}}\bar{\phi}_{1,2}) \left\{
\begin{array}{c}
X_{(5,\bar{n})} 
\\
Z_{(5,\bar{n})} 
\end{array}
\right\}
R_{zz}^{((5,1),(5,3))}(P_{\bar{\phi}}\bar{\phi}_{1,2}) |G'_{\rm proj}\rangle = \left\{
\begin{array}{c}
(-1)^{s_{(2,2)}+s_{(3,\bar{n})}}
\\
(-1)^{s_{(2,n)}+s_{(3,2)}+s_{(4,\bar{n})}}
\end{array}
\right\}
|G'_{\rm proj}\rangle.
\label{EigenvalueEq_Rzz}
\end{align}
Finally, comparison of Eq.~\eqref{EigenvalueEq_Rzz} with  Eqs.~\eqref{Eigenvalue_equation_x} and \eqref{Eigenvalue_equation_z} lets us conclude that $U_g = R_{zz}^{(1,2)}(-P_{\bar{\phi}}\bar{\phi}_{1,2}) U_{\rm swap}^{(1,2)}$ with $U_{\rm swap}^{(1,2)}$ swapping two qubits at site 1 and 2, and
\begin{align}
U_\Sigma = Z_1^{s_{(1,1)}+s_{(2,2)}+s_{(3,3)}} X_1^{s_{(2,1)}+s_{(3,2)}+s_{(4,3)}} Z_2^{s_{(1,3)}+s_{(2,2)}+s_{(3,1)}} X_2^{s_{(2,3)}+s_{(3,2)}+s_{(4,1)}},
\end{align}
where $Z_1\equiv Z_{(5,1)}$, $X_1\equiv X_{(5,1)}$, $Z_2\equiv Z_{(5,3)}$, $X_2\equiv X_{(5,3)}$. Switching $U_g$ and $U_\Sigma$, and comparing $U_g$ with Eq.~\eqref{N_qubit_rotation} ($N=2$) up to $U_{\rm swap}^{(1,2)}$, we find the measurement angle:
\begin{align}
\bar{\phi}_{1,2} = -(-1)^{s_{(2,1)}+s_{(2,3)}+s_{(3,2)}}\theta.
\end{align}
We have therefore shown by construction that the measurement pattern in the left side of Figure~1(b) 
leads to $R_{zz}^{(1,2)}(\theta)$.  To implement  
effective time evolution of a hop between sites 1 and 2, we then add Euler rotations via measurements to the end qubits in Figure~1(b) to realize a JW string [See Sec.~\ref{sub_sec_subroutines_fermions} and Figure~2(a) in the main text].

\subsection{Three-site ($N=3$) Jordan-Wigner string}

We now show that the above procedure can be generalized to a larger number of qubits by constructing a three-qubit Pauli-$z$ rotation, $R_{zzz}^{(1,2,3)}(\theta)$.  We build a SLCS with 29 qubits [see Figure~\ref{fig_coordinate_systems}(b)], defined by the composite stabilizers:
\begin{align}
K_{1,n} K_{2,m} K_{3,3} K_{4,\bar{m}} K_{5,\bar{n}} K_{7,\bar{n}},
~~
K_{2,n} K_{3,m} K_{4,3} K_{5,\bar{m}} K_{6,\bar{n}},
&~~
K_{1,3} K_{2,2} K_{2,4} K_{3,1} K_{3,3} K_{3,5} K_{4,2} K_{4,4} K_{5,3} K_{7,3},
\nonumber\\
K_{2,3} K_{3,2} K_{3,4} K_{4,1} K_{4,3} K_{4,5} K_{5,2} K_{5,4} K_{6,3},
&~~
K_{4,3} K_{5,2} K_{5,4} K_{6,1} K_{6,3} K_{6,5},
\nonumber
\end{align}
where $(n,m)=(1,2)$ if $(\bar{n},\bar{m})=(5,4)$ and vice versa. Carrying out the first round of Pauli-$x$ measurements on qubits at all sites but (3,3) in the body section, we find that the first-round projected SLCS $|G_{\rm proj}\rangle$ satisfies:
\begin{align}
Z_{(3,3)} Z_{(7,1)} Z_{(7,3)} Z_{(7,5)} |G_{\rm proj}\rangle 
& = (-1)^{s_{(4,3)}+s_{(5,2)}+s_{(5,4)}+s_{(6,1)}+s_{(6,3)}+s_{(6,5)}} |G_{\rm proj}\rangle,
\label{EigenvalueEq_N3_ZZZZ}
\\
X_{(1,n)} X_{(3,3)} X_{(7,\bar{n})} |G_{\rm proj}\rangle 
& = (-1)^{s_{(2,m)}+s_{(4,\bar{m})}+s_{(5,\bar{n})}} |G_{\rm proj}\rangle,
\label{EigenvalueEq_N3_XXX}
\\
X_{(1,3)} X_{(3,3)} X_{(7,3)} |G_{\rm proj}\rangle 
& = (-1)^{s_{(2,2)}+s_{(2,4)}+s_{(3,1)}+s_{(3,5)}+s_{(4,2)}+s_{(4,4)}+s_{(5,3)}} |G_{\rm proj}\rangle,
\\
Z_{(1,n)} Z_{(7,\bar{n})} |G_{\rm proj}\rangle 
& = (-1)^{s_{(2,n)}+s_{(3,m)}+s_{(4,3)}+s_{(5,\bar{m})}+s_{(6,\bar{n})}} |G_{\rm proj}\rangle,
\label{EigenvalueEq_N3_ZZ1}
\\
Z_{(1,3)} Z_{(7,3)} |G_{\rm proj}\rangle 
& = (-1)^{s_{(2,3)}+s_{(3,2)}+s_{(3,4)}+s_{(4,1)}+s_{(4,3)}+s_{(4,5)}+s_{(5,2)}+s_{(5,4)}+s_{(6,3)}} |G_{\rm proj}\rangle.
\label{EigenvalueEq_N3_ZZ3}
\end{align}
Then it can be shown that combination of Eqs.~\eqref{EigenvalueEq_N3_ZZZZ} and \eqref{EigenvalueEq_N3_XXX} leads to 
\begin{align}
X_{(1,n)} R_z^{(3,3)}(\bar{\phi}_{1,2,3}) X_{(3,3)} R_z^{(3,3)}(-\bar{\phi}_{1,2,3}) R_{zzz}^{((7,1),(7,3),(7,5))}(-P_{\bar{\phi}}\bar{\phi}_{1,2,3}) X_{(7,\bar{n})} R_{zzz}^{((7,1),(7,3),(7,5))}(P_{\bar{\phi}}\bar{\phi}_{1,2,3}) |G_{\rm proj}\rangle 
= Q_{\bar{\phi}} |G_{\rm proj}\rangle,
\label{EigenvalueEq_N3_XRXRRXR}
\end{align}
for arbitrary angle $\bar{\phi}_{1,2,3}$. Here we define $P_{\bar{\phi}} = (-1)^{s_{(4,3)}+s_{(5,2)}+s_{(5,4)}+s_{(6,1)}+s_{(6,3)}+s_{(6,5)}}$ and $Q_{\bar{\phi}} = (-1)^{s_{(2,m)}+s_{(4,\bar{m})}+s_{(5,\bar{n})}}$. It turns out that Eq.~\eqref{EigenvalueEq_N3_ZZ1} can be recast to become a counterpart of Eq.~\eqref{EigenvalueEq_N3_XRXRRXR}. By replacement $X_{(1,n)}\rightarrow X_{(1,3)}$, $X_{(7,\bar{n})}\rightarrow X_{(7,3)}$, and $Q_{\bar{\phi}}\rightarrow (-1)^{s_{(2,2)}+s_{(2,4)}+s_{(3,1)}+s_{(3,5)}+s_{(4,2)}+s_{(4,4)}+s_{(5,3)}}$, we can also find a similar equation to Eq.~\eqref{EigenvalueEq_N3_XRXRRXR}, that is a counterpart to Eq.~\eqref{EigenvalueEq_N3_ZZ3}. Carrying out the second round of measurement on qubit at site (3,3) in the basis $\mathcal{B}_{(3,3)}(\bar{\phi}_{1,2,3})$, we find that the second-round projected SLCS $|G'_{\rm proj}\rangle$ satisfies: 
\begin{align}
& \left\{
\begin{array}{c}
X_{(1,n)}
\\
Z_{(1,n)}
\\
X_{(1,3)}
\\
Z_{(1,3)}
\end{array}
\right\}
R_{zzz}^{((7,1),(7,3),(7,5))}(-P_{\bar{\phi}}\bar{\phi}_{1,2,3}) 
\left\{
\begin{array}{c}
X_{(7,\bar{n})}
\\
Z_{(7,\bar{n})}
\\
X_{(7,3)}
\\
Z_{(7,3)}
\end{array}
\right\}
R_{zzz}^{((7,1),(7,3),(7,5))}(P_{\bar{\phi}}\bar{\phi}_{1,2,3}) |G'_{\rm proj}\rangle 
\nonumber\\
& = 
\left\{
\begin{array}{c}
(-1)^{s_{(2,m)}+s_{(3,3)}+s_{(4,\bar{m})}+s_{(5,\bar{n})}} 
\\
(-1)^{s_{(2,n)}+s_{(3,m)}+s_{(4,3)}+s_{(5,\bar{m})}+s_{(6,\bar{n})}}
\\
(-1)^{s_{(2,2)}+s_{(2,4)}+s_{(3,1)}+s_{(3,3)}+s_{(3,5)}+s_{(4,2)}+s_{(4,4)}+s_{(5,3)}}
\\
(-1)^{s_{(2,3)}+s_{(3,2)}+s_{(3,4)}+s_{(4,1)}+s_{(4,3)}+s_{(4,5)}+s_{(5,2)}+s_{(5,4)}+s_{(6,3)}}
\end{array}
\right\}
|G'_{\rm proj}\rangle.
\label{EigenvalueEq_Rzzz}
\end{align}
Finally, comparison of Eq.~\eqref{EigenvalueEq_Rzzz} with  Eqs.~\eqref{Eigenvalue_equation_x} and \eqref{Eigenvalue_equation_z} lets us conclude that
$U_g = R_{zzz}^{(j,k,l)}(-P_{\bar{\phi}} \bar{\phi}_{1,2,3}) U_{\rm swap}^{(1,2,3)}$ with $U_{\rm swap}^{(1,2,3)}$ swapping two qubits at site 1 and 3 (2: idle), and
\begin{align}
U_\Sigma & = Z_1^{s_{(1,1)}+s_{(2,2)}+s_{(3,3)}+s_{(4,4)}+s_{(5,5)}} X_1^{s_{(2,1)}+s_{(3,2)}+s_{(4,3)}+s_{(5,4)}+s_{(6,5)}}
\nonumber\\
& ~~~ \times Z_2^{s_{(1,3)}+s_{(2,2)}+s_{(2,4)}+s_{(3,1)}+s_{(3,3)}+s_{(3,5)}+s_{(4,2)}+s_{(4,4)}+s_{(5,3)}}
\nonumber\\
& ~~~ \times X_2^{s_{(2,3)}+s_{(3,2)}+s_{(3,4)}+s_{(4,1)}+s_{(4,3)}+s_{(4,5)}+s_{(5,2)}+s_{(5,4)}+s_{(6,3)}}
\nonumber\\
& ~~~ \times Z_3^{s_{(1,5)}+s_{(2,4)}+s_{(3,3)}+s_{(4,2)}+s_{(5,1)}} X_3^{s_{(2,5)}+s_{(3,4)}+s_{(4,3)}+s_{(5,2)}+s_{(6,1)}},
\end{align}
where $Z_1\equiv Z_{(7,1)}$, $X_1\equiv X_{(7,1)}$, $Z_2\equiv Z_{(7,3)}$, $X_2\equiv X_{(7,3)}$, $Z_3\equiv Z_{(7,5)}$, $X_3\equiv X_{(7,5)}$. Switching $U_g$ and $U_\Sigma$, and comparing $U_g$ with Eq.~\eqref{N_qubit_rotation} ($N=3$) up to $U_{\rm swap}^{(1,2,3)}$, we find the measurement angle:
\begin{align}
\bar{\phi}_{1,2,3} = -(-1)^{s_{(2,1)}+s_{(2,3)}+s_{(2,5)}+s_{(4,1)}+s_{(4,5)}+s_{(5,2)}+s_{(5,4)}}\theta.
\end{align}
This concludes the construction of the measurement pattern needed to implement $R_{zzz}^{(1,2,3)}(\theta)$.  

By comparing the equations for $\bar{\phi}_{1,2,3}$ and $\bar{\phi}_{1,2}$, we therefore see how to systematically increase the length of the string with larger cluster states.  Figure~\ref{fig_coordinate_systems} also shows that only one adaptive measurement is needed to implement long strings because we can increase $N$ inductively.

\section{Compactification of square lattice cluster state}

In this section, we use the theorem on local Pauli measurement to prove that the two-qubit rotation operation implemented with the measurements on the left side of Figure~1(b) in the main text is equivalent to the operation implemented by the measurements depicted on the right side.  We start by referring to the Gottesman-Knill theorem~\cite{Gottesman99} showing that Clifford operations can be efficiently executed classically. This suggests carrying out the first round of Pauli-$x$ projectors in the body section of the SLCS in advance to map the original SLCS to the compactified form with simpler connectivity. This process is formulated in the following theorem~\cite{Hein04}: A local Pauli projector on the qubit at site $j$ in a SLCS yields a compactified cluster state (CCS) $|G_{\rm CCS}\rangle$ on the remaining qubits:
\begin{align}
\mathcal{P}_{a_j,m_j}^{(j)} |G_{\rm SLCS}\rangle 
= |m_j\rangle_{a_j}^{(j)} \otimes U_{a_j,m_j}^{(j)} |G_{\rm CCS}\rangle,
\label{local_Pauli_measurement}
\end{align}
for $a_j\in\{x,y,z\}$ and $m_j=\pm$. Here, the compactified graph $G_{\rm CCS}$ is defined by
\begin{align}
G_{\rm CCS} = \left\{
\begin{array}{ll}
G_{\rm SLCS} - \{j\}, & a_j = z
\\
\tau_{j}(G_{\rm SLCS}) - \{j\}, & a_j = y
\\
\tau_{k}(\tau_{j} \circ \tau_{k}(G_{\rm SLCS}) - \{j\}), & a_j = x
\end{array}
\right.
\end{align}
for arbitrary choice of $k \in N_{j}$ (nearest neighbors to $j$), up to the local unitaries:
\begin{align}
U_{z,\pm}^{(j)} = 
\prod_{l\in N_j} \left\{
\begin{array}{c}
\mathbb{I}_l
\\
Z_l
\end{array}
\right\}
,
~~~
U_{y,\pm}^{(j)} = \prod_{l\in N_j} \sqrt{\mp iZ_l},
~~~
U_{x,+}^{(j)} = \sqrt{iY_k} \prod_{l\in N_j \backslash (N_k \cup \{k\})} Z_l,
~~~
U_{x,-}^{(j)} = \sqrt{- iY_k} \prod_{l\in N_k \backslash (N_j \cup \{j\})}
Z_l,
\label{local_unitaries}
\end{align}
which are composed of the Clifford and non-Clifford parts. Here, $\sqrt{\pm i\sigma_a} \equiv e^{\pm i(\pi/4)\sigma_a}$ for $a\in\{x,y,z\}$.

In the above theorem, $G_{\rm CCS}$ is involved in the local complementation (LC) for the case of $a_j=x,y$. The LC of $G_{\rm SLCS}$ at a site $j$, i.e., $\tau_j: G_{\rm SLCS} \mapsto \tau_j(G_{\rm SLCS})$, is obtained by complementing the subgraph of $G_{\rm SLCS}$ induced by $N_j$ (i.e., disconnecting/connecting qubits belonging to $N_j$ if they are originally connected/disconnected) and leaving the rest of parts unchanged. The corresponding LC-equivalent SLCS is defined by $|\tau_j(G_{\rm SLCS})\rangle = U_\tau^{(j)}|G_{\rm SLCS}\rangle$ up to the local Clifford unitary $U_\tau^{(j)} = \sqrt{-iX_j} \prod_{l\in N_j} \sqrt{iZ_l}$.

The process to reach the CCS involves a sequence of projectors. In successive applications of Eq.~\eqref{local_Pauli_measurement}, we need to deal with Pauli propagation of the projector $\mathcal{P}_{a_j,m}^{(j)}$, because local unitaries always intervene. We note that Pauli propagation is governed by a different rule for the Clifford and non-Clifford gates. For the Clifford gates, Pauli propagation at most flips the sign of projector outcomes:
$\mathcal{P}_{x,\pm} Z = Z \mathcal{P}_{x,\mp}$, 
$\mathcal{P}_{y,\pm} Z = Z \mathcal{P}_{y,\mp}$. On the other hand, for the non-Clifford gates, it makes a more drastic impact, i.e., the reorientation of directions:
\begin{align}
\mathcal{P}_{x,\pm} \sqrt{iY} = \sqrt{iY} \mathcal{P}_{z,\pm},
~~
\mathcal{P}_{x,\pm} \sqrt{-iY} & = \sqrt{-iY} \mathcal{P}_{z,\pm},
~~
\mathcal{P}_{x,\pm} \sqrt{iZ} = \sqrt{iZ} \mathcal{P}_{y,\pm},
~~
\mathcal{P}_{x,\pm} \sqrt{-iZ} = \sqrt{-iZ} \mathcal{P}_{y,\mp},
\\
\mathcal{P}_{y,\pm} \sqrt{iZ} = \sqrt{iZ} \mathcal{P}_{x,\mp},
~~
\mathcal{P}_{y,\pm} \sqrt{-iZ} & = \sqrt{-iZ} \mathcal{P}_{x,\pm},
~~
\mathcal{P}_{z,\pm} \sqrt{iY} = \sqrt{iY} \mathcal{P}_{x,\pm},
~~
\mathcal{P}_{z,\pm} \sqrt{-iY} = \sqrt{-iY} \mathcal{P}_{x,\pm}.
\end{align}

\begin{figure}[t]
\begin{center}
\includegraphics[width=0.98\textwidth]{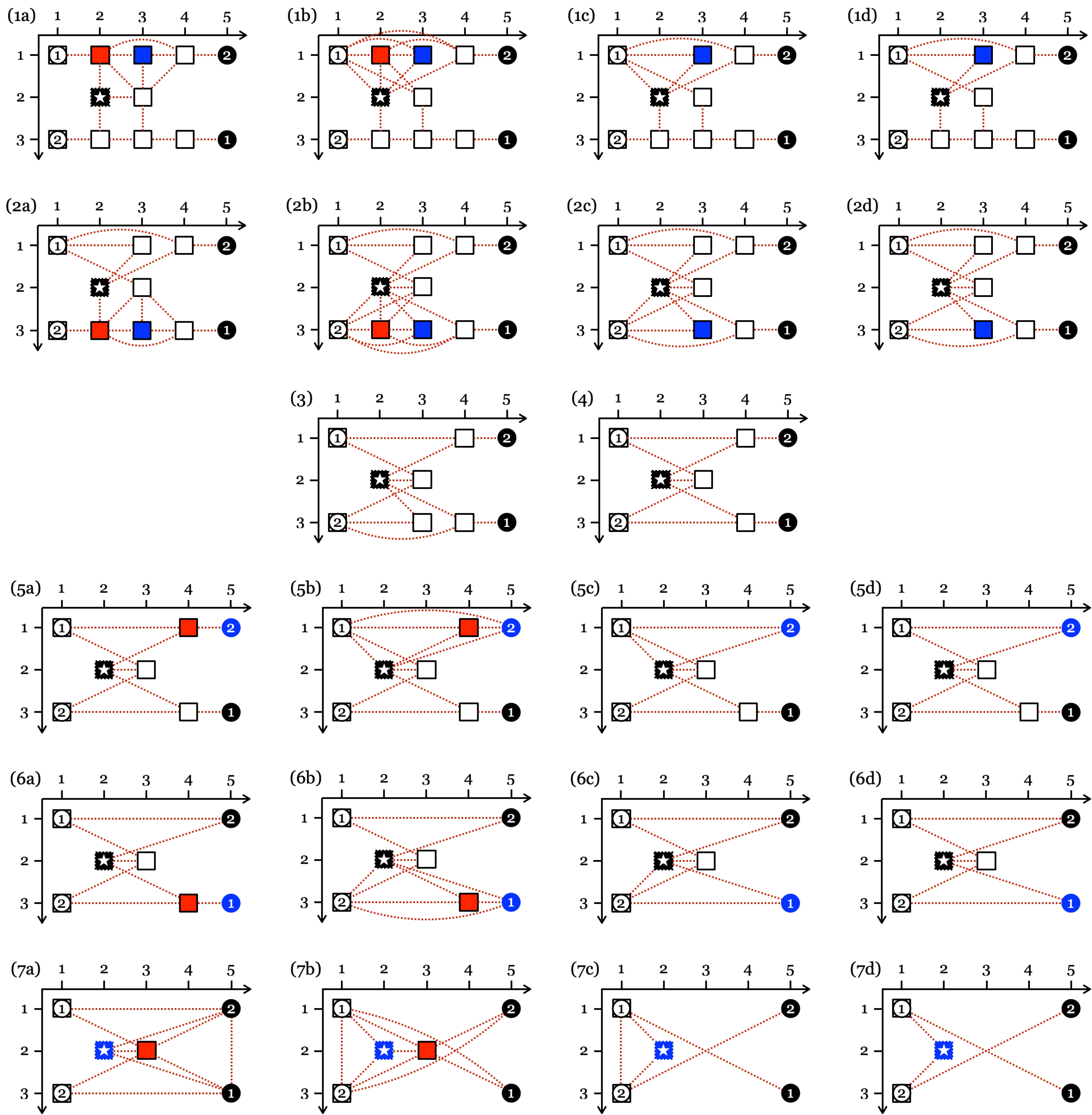}
\end{center}
\caption{Procedure to map the SLCS to the CCS for an example of 2-qubit rotation gate.}
\label{fig_ccs_steps}
\end{figure}

We now revisit the measurement pattern for the 2-qubit rotation gate [see Figure~\ref{fig_coordinate_systems}(a) or the left side of Figure~1(b) in the main text]. To find the CCS [shown in the right side of Figure~1(b) in the main text], we use the above theorem to project qubits at sites $j_1=(2,1)$, $j_2=(2,3)$, $j_3=(3,1)$, $j_4=(3,3)$, $j_5=(4,1)$, $j_6=(4,3)$, $j_7=(3,2)$. We take the following steps as summarized in Figure~\ref{fig_ccs_steps}:

{\bf Step 1:} Pauli-$x$ projector is applied at site $j_1 = (2,1)$ (red box) with special neighbor $k_1 = (3,1)$ (blue box) to find $G_{\rm CCS}^{(1)} = \tau_{k_1}(\tau_{j_1}\circ\tau_{k_1}(G_{\rm SLCS}) - \{j_1\})$ up to $U_{x,m_1}^{(j_1)}$. 
(1a) LC of $G_{\rm SLCS}$ at $k_1$, 
(1b) LC of (1a) at $j_1$, 
(1c) exclusion of $j_1$ from (1b), and 
(1d) LC of (1c) at $k_1$.

{\bf Step 2:} A Pauli-$x$ projector is applied at site $j_2 = (2,3)$ (red box) with special neighbor $k_2 = (3,3)$ (blue box) to find $G_{\rm CCS}^{(2)} = \tau_{k_2}(\tau_{j_2}\circ\tau_{k_2}(G_{\rm CCS}^{(1)}) - \{j_2\})$ up to $U_{x,m_2}^{(j_2)}$. 
(2a) LC of $G_{\rm CCS}^{(1)}$ at $k_2$, 
(2b) LC of (2a) at $j_2$, 
(2c) exclusion of $j_2$ from (2b), and 
(2d) LC of (2c) at $k_2$.

{\bf Step 3:} A Pauli-$x$ projector is applied at site $j_3 = (3,1)$ in $G_{\rm CCS}^{(2)}$. Since $\mathcal{P}_{x,m_3}^{(j_3)}$ does not commute with $\sqrt{m_1 iY_{k_1(=j_3)}}$ (obtained from Step 1), Pauli propagation is used to obtain: $\mathcal{P}_{x,m_3}^{(j_3)} \sqrt{m_1iY_{j_3}}|G_{\rm CCS}^{(2)}\rangle
= \sqrt{m_1iY_{j_3}} \mathcal{P}_{z,m_3}^{(j_3)}|G_{\rm CCS}^{(2)}\rangle$. Thus Pauli-$z$ measurement is effectively carried out at $j_3$, leading to the exclusion of $j_3$.

{\bf Step 4:} A Pauli-$x$ projector is applied at site $j_4 = (3,3)$ in $G_{\rm CCS}^{(3)}$. Since $\mathcal{P}_{x,m_4}^{(j_4)}$ does not commute with $\sqrt{m_2 iY_{k_2(=j_4)}}$ (obtained from Step 2), Pauli propagation is used to obtain: $\mathcal{P}_{x,m_4}^{(j_4)} \sqrt{m_2iY_{j_4}}|G_{\rm CCS}^{(3)}\rangle
= \sqrt{m_2iY_{j_4}} \mathcal{P}_{z,m_4}^{(j_4)}|G_{\rm CCS}^{(3)}\rangle$. Thus Pauli-$z$ measurement is effectively carried out at $j_4$, leading to the exclusion of $j_4$.

{\bf Step 5:} A Pauli-$x$ projector is applied at site $j_5 = (4,1)$ (red box) with special neighbor $k_5 = (5,1)$ (blue box) to find $G_{\rm CCS}^{(5)} = \tau_{k_5}(\tau_{j_5}\circ\tau_{k_5}(G_{\rm CCS}^{(4)}) - \{j_5\})$ up to $U_{x,m_5}^{(j_5)}$. 
(5a) LC of $G_{\rm CCS}^{(4)}$ at $k_5$, 
(5b) LC of (5a) at $j_5$, 
(5c) exclusion of $j_5$ from (5b), and 
(5d) LC of (5c) at $k_5$.

{\bf Step 6:} A Pauli-$x$ projector is applied at site $j_6 = (4,3)$ (red box) with special neighbor $k_6 = (5,3)$ (blue box) to find $G_{\rm CCS}^{(6)} = \tau_{k_6}(\tau_{j_6}\circ\tau_{k_6}(G_{\rm CCS}^{(5)}) - \{j_6\})$ up to $U_{x,m_6}^{(j_6)}$. 
(6a) LC of $G_{\rm CCS}^{(5)}$ at $k_6$, 
(6b) LC of (6a) at $j_6$, 
(6c) exclusion of $j_6$ from (6b), and 
(6d) LC of (6c) at $k_6$.

{\bf Step 7:} A Pauli-$x$ projector is applied at site $j_7 = (3,2)$ (red box) with special neighbor $k_7 = (2,2)$ (blue box) to find $G_{\rm CCS} = \tau_{k_7}(\tau_{j_7}\circ\tau_{k_7}(G_{\rm CCS}^{(6)}) - \{j_7\})$ up to $U_{x,m_7}^{(j_7)}$. 
(7a) LC of $G_{\rm CCS}^{(6)}$ at $k_7$, 
(7b) LC of (7a) at $j_7$, 
(7c) exclusion of $j_7$ from (7b), and 
(7d) LC of (7c) at $k_7$.

We note that the final result $G_{\rm CCS}$ depends on the choice of special neighbors ($k_1$, $k_2$, $k_5$, $k_6$, $k_7$). But it turns out that any variation belongs to the same LC-equivalent class up to local Clifford unitaries. This completes the constructive proof showing the equivalence of the results of measurements depicted on left and right sides of Figure~1(b) in the main text.

\section{MBQC subroutines for fermion systems}
\label{sub_sec_subroutines_fermions}

In this section, we derive two expressions stated in the main text: (1) the signs for the measurement angles, $S_\theta^\textrm{K}$ and $S_\theta^\textrm{H}$, and (2) the byproduct operators of the MBQC subroutines for the Kitaev and Hubbard chains. We also prove that, as stated in the main text, the measurement pattern for the Kitaev and Hubbard chains can be concatenated to time-evolve larger $N$ or $M$.

\subsection{Kitaev chain}

We start with the first-order Trotterized form of $e^{-iH_\textrm{K}t}$ [Eq.~(2) in the main text]:
\begin{align}
U_g & = \prod_{j=1}^{N-1} R_{xx}^{(j,j+1)}(-2\phi_M) 
\prod_{k=1}^{N} R_z^{(k)}(-2g_\mu\phi_M).
\label{Quantum_gate_Kitaev}
\end{align}
It is convenient to recast Eq.~\eqref{Quantum_gate_Kitaev} into the MBQC-adaptive form by decomposing the two-qubit rotation gate into
\begin{align}
R_{xx}^{(j,k)}(\theta) 
& = R_y^{(j)}(-\lambda) R_y^{(k)}(-\lambda) R_{zz}^{(j,k)}(\theta) R_y^{(j)}(\lambda) R_y^{(k)}(\lambda),
\end{align}
in conjunction with the Euler decomposition $R_y(\lambda) = R_x(\gamma) R_z(\beta) R_x(\alpha)$ where $-\lambda = -\alpha = \beta = \gamma = \pi/2$. The number of single-qubit rotation gates can be reduced by applying Pauli propagation to the array of gates: $R_x(\gamma) R_z(\beta) R_x(\alpha) R_z(-2g_\mu\phi_M)$ $= R_x(2g_\mu\phi_M+\gamma) R_z(\beta) R_x(\alpha)$. After some algebra, for $N=2$, Eq.~\eqref{Quantum_gate_Kitaev} has the form:
\begin{align}
U_g & = R_x^{(1)}(-\alpha) R_z^{(1)}(-\beta) R_x^{(1)}(-\gamma) R_x^{(2)}(-\alpha) R_z^{(2)}(-\beta) R_x^{(2)}(-\gamma) R_{zz}^{(1,2)}(-2\phi_M)
\nonumber\\
& ~~~ \times R_x^{(1)}(2g_\mu\phi_M+\gamma) R_z^{(1)}(\beta) R_x^{(1)}(\alpha) R_x^{(2)}(2g_\mu\phi_M+\gamma) R_z^{(2)}(\beta) R_x^{(2)}(\alpha).
\label{Gate_Kitaev_N2}
\end{align}
For larger $N$, Eq.~\eqref{Gate_Kitaev_N2} can be concatenated in the following way:
\begin{align}
U_g = \left\{
\begin{array}{cl}
\tilde{W}_g^{(N-1)} W_g^{(1)},
& N = 3,
\\
\tilde{W}_g^{(N-1)} \Big[\prod_{j=2}^{N-2}V_g^{(j)}\Big] W_g^{(1)},
& N \geq 4,
\end{array}
\right.
\label{Gate_Kitaev_N3}
\end{align}
where we define three types of composite gates:
\begin{align}
W_g^{(1)} & = R_x^{(1)}(-\alpha) R_z^{(1)}(-\beta) R_x^{(1)}(-\gamma) R_{zz}^{(1,2)}(-2\phi_M)
R_x^{(1)}(2g_\mu\phi_M+\gamma) R_z^{(1)}(\beta) R_x^{(1)}(\alpha)
\nonumber\\
& ~~~ \times R_x^{(2)}(2g_\mu\phi_M+\gamma) R_z^{(2)}(\beta) R_x^{(2)}(\alpha),
\label{Gate_Kitaev_W1}
\\
V_g^{(j)} & = R_x^{(j)}(-\alpha) R_z^{(j)}(-\beta) R_x^{(j)}(-\gamma) R_{zz}^{(j,j+1)}(-2\phi_M) R_x^{(j+1)}(2g_\mu\phi_M+\gamma) R_z^{(j+1)}(\beta) R_x^{(j+1)}(\alpha),
\label{Gate_Kitaev_V}
\\
& ~~~~~(2\leq j\leq N-2)~(N\geq 4)
\nonumber
\end{align}
\begin{align}
\tilde{W}_g^{(N-1)} & = R_x^{(N-1)}(-\alpha) R_z^{(N-1)}(-\beta) R_x^{(N-1)}(-\gamma)
R_x^{(N)}(-\alpha) R_z^{(N)}(-\beta) R_x^{(N)}(-\gamma)
R_{zz}^{(N-1,N)}(-2\phi_M)
\nonumber\\
& ~~~ \times R_x^{(N)}(2g_\mu\phi_M+\gamma) R_z^{(N)}(\beta) R_x^{(N)}(\alpha).
~~
(N\geq 3)
\label{Gate_Kitaev_WN}
\end{align}

\begin{figure}[t]
\begin{center}
\includegraphics[width=0.84\textwidth]{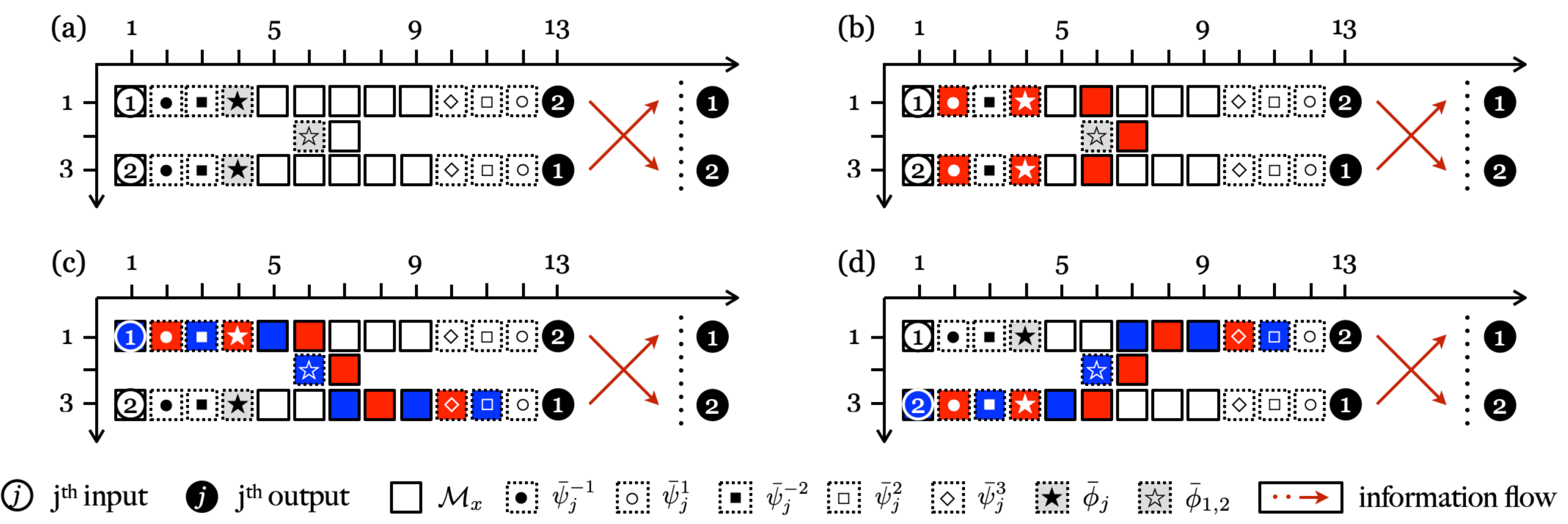}
\end{center}
\caption{(a) Coordinate systems for assigning qubit positions in the measurement pattern for the Kitaev chain with $N=2$. In panel (b-d), we highlight correlation centers contributing to $S_{\bar{\phi}_{1,2}}$ [Eq.~\eqref{S_kitaev_12}; red boxes in (b)], $\{S_{\bar{\psi}^{-1}_1}, S_{\bar{\phi}_1}, S_{\bar{\psi}^3_1}, S_{\bar{\psi}^1_1}\}$ [Eq.~\eqref{S_kitaev_group_1}; blue boxes in (c)], $\{S_{\bar{\psi}^{-2}_1}, S_{\bar{\psi}^2_1}\}$ [Eq.~\eqref{S_kitaev_group_2}; red boxes in (c)], $\{S_{\bar{\psi}^{-1}_2}, S_{\bar{\phi}_2}, S_{\bar{\psi}^3_2}, S_{\bar{\psi}^1_2}\}$ [Eq.~\eqref{S_kitaev_group_3}; blue boxes in (d)], and $\{S_{\bar{\psi}^{-2}_2}, S_{\bar{\psi}^2_2}\}$ [Eq.~\eqref{S_kitaev_group_4}; red boxes in (d)].}
\label{fig_subroutine_kitaev_N2}
\end{figure}

Before dealing with large $N$, we first consider the simplest example with $N=2$. The measurement pattern for implementing $U_g$ ($N=2$) is shown in Figure~\ref{fig_subroutine_kitaev_N2}(a).   The measurement pattern is composed of 5 parts: the main SLCS for a two-qubit rotation gate and 4 legs for Euler-decomposed single-qubit rotation gates. To justify this, we do not repeat the analysis in the stabilizer formalism, but instead combine the known results for 5 different gates (see Sec.~\ref{sec_measurement_pattern_JW_string}). We start with the following array of unitary gates and byproduct operators:
\begin{align}
& U_{\rm swap}^{(1,2)} \big[R_x^{(1)}(-\bar{\psi}_{1}^1) X_{1}^{s_{(12,1)}} R_z^{(1)}(-\bar{\psi}^2_{1}) Z_{1}^{s_{(11,1)}} R_x^{(1)}(-\bar{\psi}^3_{1}) Z_{1}^{s_{(9,1)}} X_{1}^{s_{(10,1)}}\big]
\nonumber\\
& \times \big[R_x^{(2)}(-\bar{\psi}^1_{2}) X_{2}^{s_{(12,3)}} R_z^{(2)}(-\bar{\psi}^2_{2}) Z_{2}^{s_{(11,3)}} R_x^{(2)}(-\bar{\psi}^3_{2}) Z_{2}^{s_{(9,3)}} X_{2}^{s_{(10,3)}}\big]
\nonumber\\
& \times \big[R_{zz}^{(1,2)}(-(-1)^{s_{(7,2)}+s_{(8,1)}+s_{(8,3)}}\bar{\phi}_{1,2}) U_{\rm swap}^{(1,2)} Z_{1}^{s_{(5,1)}+s_{(6,2)}+s_{(7,3)}} X_{1}^{s_{(6,1)}+s_{(7,2)}+s_{(8,3)}} Z_{2}^{s_{(5,3)}+s_{(6,2)}+s_{(7,1)}} X_{2}^{s_{(6,3)}+s_{(7,2)}+s_{(8,1)}}\big] 
\nonumber\\
& \times \big[R_x^{(1)}(-\bar{\phi}_{1}) X_{1}^{s_{(4,1)}} R_z^{(1)}(-\bar{\psi}^{-2}_{1}) Z_{1}^{s_{(3,1)}} R_x^{(1)}(-\bar{\psi}^{-1}_{1}) Z_{1}^{s_{(1,1)}} X_{1}^{s_{(2,1)}}\big]
\nonumber\\
& \times \big[R_x^{(2)}(-\bar{\phi}_{2}) X_{2}^{s_{(4,3)}} R_z^{(2)}(-\bar{\psi}^{-2}_{2}) Z_{2}^{s_{(3,3)}} R_x^{(2)}(-\bar{\psi}^{-1}_{2}) Z_{2}^{s_{(1,3)}} X_{2}^{s_{(2,3)}}\big],
\label{Gate_Kitaev_combined_N_2}
\end{align}
where the expression in the third bracket comes from the two-qubit rotation gate, while the remaining expressions come from the Euler-decomposed single-qubit rotation gates. The next step is to push all byproduct operators to the left side of all other unitary gates by successively applying Pauli propagation (Here, two swap gates are canceled). Then Eq.~\eqref{Gate_Kitaev_combined_N_2} is recast into the form: $U_\Sigma \tilde{U}_g$, where $U_\Sigma$ is the total byproduct operator. Matching $\tilde{U}_g$ with $U_g$ [Eq.~\eqref{Gate_Kitaev_N2}], we find the expressions for measurement angles:
\begin{align}
\bar{\phi}_{1,2} = 2P_{\bar{\phi}_{1,2}} \phi_M,
~~
\bar{\phi}_{j} = -P_{\bar{\phi}_j} (2g_\mu\phi_M + \gamma),
~~
\bar{\psi}_{j}^{r} =  P_{\bar{\psi}_{j}^{r}} \psi^{r},
\label{Measurement_angle_Kitaev_N2}
\end{align}
where $\psi^r\in\{\pm\alpha,\pm\beta,\gamma\}$ for $r=\pm1,\pm2,3$, and $P_{\theta} = (-1)^{S_\theta}$. The exponent $S_\theta$ accumulates all measurement outcomes during single-qubit measurements. Specifically, the exponent for the two-qubit rotation gate is defined by [see Figure~\ref{fig_subroutine_kitaev_N2}(b)]
\begin{align}
S_{\bar{\phi}_{1,2}} & = s_{(2,1)}+s_{(2,3)}+s_{(4,1)}+s_{(4,3)}+s_{(6,1)}+s_{(6,3)}+s_{(7,2)}.
\label{S_kitaev_12}
\end{align}
Other exponents for the single-qubit rotation gates are defined in the hierarchical form [see Figure~\ref{fig_subroutine_kitaev_N2}(c-d)]:
\begin{align}
& \left\{
\begin{array}{rl}
S_{\bar{\psi}^{-1}_1} 
& = s_{(1,1)},
\\
S_{\bar{\phi}_1}
& = S_{\bar{\psi}^{-1}_1}+s_{(3,1)},
\\
S_{\bar{\psi}^3_1} 
& = S_{\bar{\phi}_1}+s_{(5,1)}+s_{(6,2)}+s_{(7,3)}+s_{(9,3)},
\\
S_{\bar{\psi}^1_1} 
& = S_{\bar{\psi}^3_1}+s_{(11,3)},
\end{array}
\right.
\label{S_kitaev_group_1}
\\
& \left\{
\begin{array}{rl}
S_{\bar{\psi}^{-2}_1} 
& = s_{(2,1)},
\\
S_{\bar{\psi}^2_1} 
& = S_{\bar{\psi}^{-2}_1}+s_{(4,1)}+s_{(6,1)}+s_{(7,2)}+s_{(8,3)}+s_{(10,3)},
\end{array}
\right.
\label{S_kitaev_group_2}
\end{align}
\begin{align}
& \left\{
\begin{array}{rl}
S_{\bar{\psi}^{-1}_2} 
& = s_{(1,3)},
\\
S_{\bar{\phi}_2} 
& = S_{\bar{\psi}^{-1}_2}+s_{(3,3)},
\\
S_{\bar{\psi}^3_2} 
& = S_{\bar{\phi}_2}+s_{(5,3)}+s_{(6,2)}+s_{(7,1)}+s_{(9,1)},
\\
S_{\bar{\psi}^1_2} 
& = S_{\bar{\psi}^3_2}+s_{(11,1)},
\end{array}
\right.
\label{S_kitaev_group_3}
\\
& \left\{
\begin{array}{rl}
S_{\bar{\psi}^{-2}_2} 
& = s_{(2,3)},
\\
S_{\bar{\psi}^2_2} 
& = S_{\bar{\psi}^{-2}_2}+s_{(4,3)}+s_{(6,3)}+s_{(7,2)}+s_{(8,1)}+s_{(10,1)}.
\end{array}
\right.
\label{S_kitaev_group_4}
\end{align}
We also find the total byproduct operator:
\begin{align}
U_\Sigma = Z_1^{S_{Z_1}} X_1^{S_{X_1}} Z_2^{S_{Z_2}} X_2^{S_{X_2}},
\label{Byproduct_Kitaev}
\end{align}
where the exponents are defined by using Eqs.~\eqref{S_kitaev_group_1}-\eqref{S_kitaev_group_4}:
\begin{align}
S_{Z_1} = S_{\bar{\psi}^1_1},
~~
S_{X_1} = S_{\bar{\psi}^2_1}+s_{(12,3)},
~~
S_{Z_2} = S_{\bar{\psi}^1_2},
~~
S_{X_2} = S_{\bar{\psi}^2_2}+s_{(12,1)}.
\label{S_kitaev_byproduct}
\end{align}

\begin{figure}[t]
\begin{center}
\includegraphics[width=0.84\textwidth]{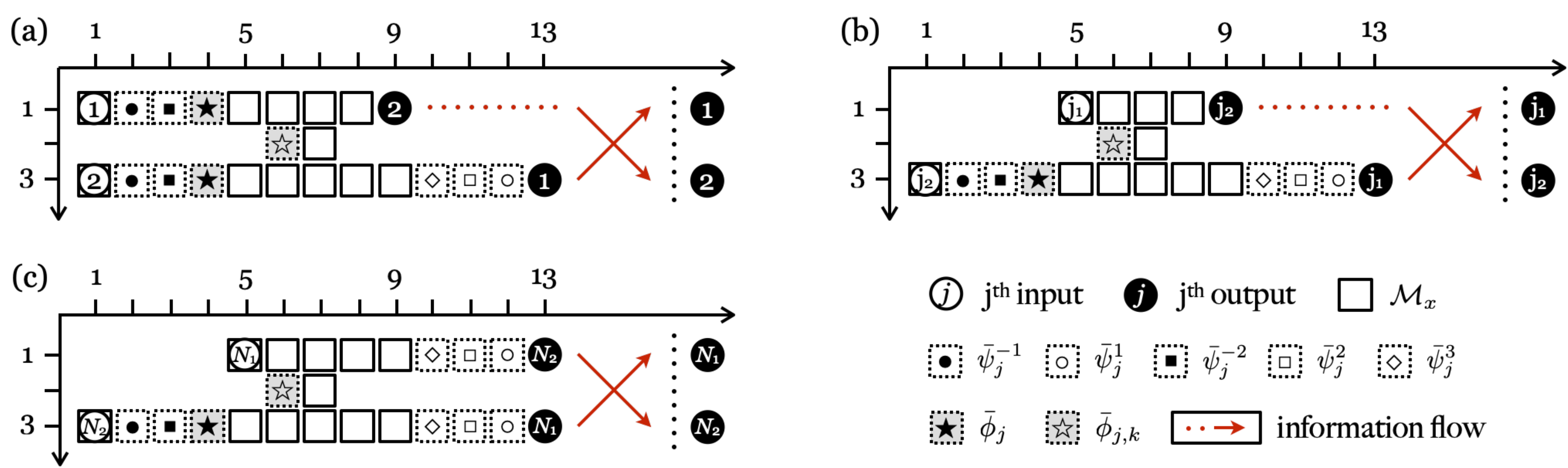}
\end{center}
\caption{Coordinate systems for assigning qubit positions in the giant measurement pattern for the Kitaev chain with $N\geq 3$, which can be built by combining three types of measurement patterns for (a) $W_g^{(1)}$, (b) $V_g^{(j)}$ ($2\leq j\leq N-2) (N\geq 4$), and (c) $\tilde{W}_g^{(N-1)}$ ($N\geq 3$). Here, input and output qubits are indexed by $(j_1, j_2) = (j, j+1)$ for a given $j$ in (b), and $(N_1, N_2) = (N-1, N)$ for a given $N$ in (c).}
\label{fig_subroutine_kitaev_N3}
\end{figure}

We now concatenate the $N=2$ result to $N\geq 3$. We proceed in two steps: First, we build three types of measurement patterns for $W_g^{(1)}$, $V_g^{(j)}$, $\tilde{W}_g^{(N-1)}$ [Eqs.~\eqref{Gate_Kitaev_W1}-\eqref{Gate_Kitaev_WN}] (Figure~\ref{fig_subroutine_kitaev_N3}). Second, we combine them in a specific order for a given $N$. The resulting giant measurement pattern has a cascade structure flowing from the left top to the right bottom. It turns out that there is no change in Eq.~\eqref{Measurement_angle_Kitaev_N2} but with generalization:
\begin{align}
\bar{\phi}_{j,k} = 2P_{\bar{\phi}_{j,k}} \phi_M.
\label{Measurement_angle_Kitaev_N3}
\end{align}
The exponents in Eqs.~\eqref{S_kitaev_12}-\eqref{S_kitaev_group_4} are modified into:
\begin{align}
S^{(j)}_{\bar{\phi}_{j,j+1}} = \big[s_{(2,1)}^{(1)}+s_{(4,1)}^{(1)}\big] \delta_{j,1}+s_{(2,3)}^{(j)}+s_{(4,3)}^{(j)}+s_{(6,1)}^{(j)}+s_{(6,3)}^{(j)}+s_{(7,2)}^{(j)},
\end{align}
and
\begin{align}
& \left\{
\begin{array}{rl}
S^{(j)}_{\bar{\psi}^{-1}_{j}} 
& = s_{(1,1)}^{(1)}\delta_{j,1},
\\
S^{(j)}_{\bar{\phi}_{j}} 
& = S^{(j)}_{\bar{\psi}^{-1}_{j}}+s_{3,1}^{(1)}\delta_{j,1},
\\
S^{(j)}_{\bar{\psi}^3_{j}} 
& = S^{(j)}_{\bar{\phi}_{j}}+s_{(5,1)}^{(j)}+s_{(6,2)}^{(j)}+s_{(7,3)}^{(j)}+s_{(9,3)}^{(j)},
\\
S^{(j)}_{\bar{\psi}^1_{j}} 
& = S^{(j)}_{\bar{\psi}^3_{j}}+s_{(11,3)}^{(j)},
\end{array}
\right.
\\
& \left\{
\begin{array}{rl}
S^{(j)}_{\bar{\psi}^{-2}_{j}} 
& = s_{(2,1)}^{(1)}\delta_{j,1},
\\
S^{(j)}_{\bar{\psi}^2_{j}} 
& = S^{(j)}_{\bar{\psi}^{-2}_{j}}+s_{(4,1)}^{(1)}\delta_{j,1}+ s_{(6,1)}^{(j)}+s_{(7,2)}^{(j)}+s_{(8,3)}^{(j)}+s_{(10,3)}^{(j)},
\end{array}
\right.
\end{align}
\begin{align}
& \left\{
\begin{array}{rl}
S^{(j)}_{\bar{\psi}^{-1}_{j+1}} 
& = s_{(1,3)}^{(j)},
\\
S^{(j)}_{\bar{\phi}_{j+1}} 
& = S^{(j)}_{\bar{\psi}^{-1}_{j+1}}+s_{(3,3)}^{(j)},
\\
S^{(j)}_{\bar{\psi}^3_{j+1}} 
& = S^{(j)}_{\bar{\phi}_{j+1}}+s_{(5,3)}^{(j)}+s_{(6,2)}^{(j)}+s_{(7,1)}^{(j)}+s_{(9,1)}^{(N-1)}\delta_{j,N-1},
\\
S^{(j)}_{\bar{\psi}^1_{j+1}} 
& = S^{(j)}_{\bar{\psi}^3_{j+1}}+s_{(11,1)}^{(N-1)}\delta_{j,N-1},
\end{array}
\right.
\\
& \left\{
\begin{array}{rl}
S^{(j)}_{\bar{\psi}^{-2}_{j+1}} 
& = s_{(2,3)}^{(j)},
\\
S^{(j)}_{\bar{\psi}^2_{j+1}} 
& = S^{(j)}_{\bar{\psi}^{-2}_{j+1}}+s_{(4,3)}^{(j)}+ s_{(6,3)}^{(j)}+s_{(7,2)}^{(j)}+s_{(8,1)}^{(j)}+s_{(10,1)}^{(N-1)}\delta_{j,N-1},
\end{array}
\right.
\end{align}
respectively.  Here the superscripts in $s^{(j)}$ and $S^{(j)}$ are to be consistent with $W_g^{(1)}$, $V_g^{(j)}$, $\tilde{W}_g^{(N-1)}$, and $\delta_{j,j'}$ is the Kronecker delta. The exponents in the byproduct operator have forms similar to Eq.~\eqref{S_kitaev_byproduct}:
\begin{align}
S^{(j)}_{Z_{j}} = S^{(j)}_{\bar{\psi}^1_j},
~~
S^{(j)}_{X_{j}} = S^{(j)}_{\bar{\psi}^2_j}+s_{(12,3)}^{(j)},
~~
S^{(j)}_{Z_{j+1}} = S^{(j)}_{\bar{\psi}^1_{j+1}},
~~
S^{(j)}_{X_{j+1}} = S^{(j)}_{\bar{\psi}^2_{j+1}}+s_{(12,1)}^{(N-1)}\delta_{j,N-1}.
\end{align}
To complete the concatenation process, we need to reapply Pauli propagation to push all byproduct operators to the left side of all other unitary gates in the giant measurement pattern. In this process, the exponents in $V_g^{(j)}$ are adjusted to accumulate further the measurement outcomes in $V_g^{(j-1)}$. The same thing also happens for the pairs: $\{V_g^{(2)}, W_g^{(1)}\}$ and $\{\tilde{W}_g^{(N-1)}, V_g^{(N-2)}\}$. Consequently, some exponents are modified in the following way:
\begin{align}
S^{(j)}_{\bar{\psi}^1_j} 
\rightarrow S^{(j)}_{\bar{\psi}^1_j} + S^{(j-1)}_{Z_{j}},
~~
S^{(j)}_{\bar{\psi}^2_j} 
\rightarrow S^{(j)}_{\bar{\psi}^2_j} + S^{(j-1)}_{X_{j}},
~~
S^{(j)}_{\bar{\psi}^3_j} 
\rightarrow S^{(j)}_{\bar{\psi}^3_j} + S^{(j-1)}_{Z_{j}},
~~
S^{(j)}_{\bar{\phi}_{j,j+1}} 
\rightarrow S^{(j)}_{\bar{\phi}_{j,j+1}} + S^{(j-1)}_{X_{j}}.
\end{align}

This concludes our derivation of expressions discussed in the main text:  (1) the signs for the measurement angles, $S_\theta^\textrm{K}$, and (2) the byproduct operators for the Kitaev chain.  We have also proven the statement that the measurement pattern for the Kitaev chain shown in the main text can be concatenated for larger chains.

\subsection{Hubbard chain}

We start with the first-order Trotterized form of $e^{-iH_\textrm{H}t}$ [Eq.~(6) in the main text]:
\begin{align}
U_g = \prod_{j=1}^{N} \big[ R_{zz}^{(2j-1,2j)}(g_U\phi_M) R_z^{(2j-1)}(g_U\phi_M) R_z^{(2j)}(g_U\phi_M) \big] \prod_{k=1}^{2N-2} \big[ R_{xzx}^{(k,k+1,k+2)}(\phi_M) R_{yzy}^{(k,k+1,k+2)}(\phi_M) \big].
\label{Quantum_gate_Hubbard}
\end{align}
It is convenient to recast Eq.~\eqref{Quantum_gate_Hubbard} into the MBQC-adaptive form by decomposing the three-qubit rotation gates into
\begin{align}
R_{xzx}^{(j,k,l)}(\theta) 
& = R_y^{(j)}(-\lambda) R_y^{(l)}(-\lambda) R_{zzz}^{(j,k,l)}(\theta) R_y^{(j)}(\lambda) R_y^{(l)}(\lambda),
\\
R_{yzy}^{(j,k,l)}(\theta) 
& = R_x^{(j)}(\lambda) R_x^{(l)}(\lambda) R_{zzz}^{(j,k,l)}(\theta) R_x^{(j)}(-\lambda) R_x^{(l)}(-\lambda),
\end{align}
in conjunction with the Euler decomposition $R_y(\lambda) = R_x(\gamma) R_z(\beta) R_x(\alpha)$ where $-\lambda = -\alpha = \beta = \gamma = \pi/2$. The number of single-qubit rotation gates can be reduced by applying Pauli propagation to the array of gates: $R_z(g_U\phi_M) R_x(-\alpha) R_z(-\beta) R_x(-\gamma)$ $= R_x(-\alpha) R_z(-\beta) R_x(-g_U\phi_M-\gamma)$. After some algebra, Eq.~\eqref{Quantum_gate_Hubbard} is rearranged into a form for use in concatenation of longer Hubbard chains:
\begin{align}
U_g = \left\{
\begin{array}{cc}
W_g^{(1)}, & N = 2
\\
W_g^{(N-1)} \prod_{j=1}^{N-2} V_g^{(j)}, & N \geq 3
\end{array}
\right.
\label{Gate_Hubbard_N}
\end{align}
where we define two types of composite gates:
\begin{align}
V_g^{(j)}
& = R_{zz}^{(2j-1,2j)}(g_U\phi_M) R_z^{(2j-1)}(g_U\phi_M)  R_x^{(2j)}(-\alpha) R_z^{(2j)}(-\beta) R_x^{(2j)}(-g_U\phi_M-\gamma) R_x^{(2j+2)}(-\alpha) R_z^{(2j+2)}(-\beta) R_x^{(2j+2)}(-\gamma) 
\nonumber\\
& ~~~ \times R_{zzz}^{(2j,2j+1,2j+2)}(\phi_M) R_x^{(2j)}(\gamma) R_z^{(2j)}(\beta) R_x^{(2j)}(\lambda+\alpha)  R_x^{(2j+2)}(\gamma) R_z^{(2j+2)}(\beta) R_x^{(2j+2)}(\lambda+\alpha)
\nonumber\\
&~~~ \times R_{zzz}^{(2j,2j+1,2j+2)}(\phi_M)   R_x^{(2j-1)}(\alpha) R_x^{(2j)}(-\alpha) R_x^{(2j+1)}(\alpha) 
R_x^{(2j+2)}(-\alpha)
\nonumber\\
& ~~~ \times R_{zzz}^{(2j-1,2j,2j+1)}(\phi_M) R_x^{(2j-1)}(-\lambda-\alpha) R_z^{(2j-1)}(-\beta) R_x^{(2j-1)}(-\gamma) R_x^{(2j+1)}(-\lambda-\alpha) R_z^{(2j+1)}(-\beta) R_x^{(2j+1)}(-\gamma)
\nonumber\\
& ~~~ \times R_{zzz}^{(2j-1,2j,2j+1)}(\phi_M) R_x^{(2j-1)}(\gamma) R_z^{(2j-1)}(\beta) R_x^{(2j-1)}(\alpha) R_x^{(2j+1)}(\gamma) R_z^{(2j+1)}(\beta) R_x^{(2j+1)}(\alpha),
\label{Gate_Hubbard_V}
\end{align}
\begin{align}
W_g^{(N-1)}
& = R_{zz}^{(2N-3,2N-2)}(g_U\phi_M) R_z^{(2N-3)}(g_U\phi_M) R_x^{(2N-2)}(-\alpha) R_z^{(2N-2)}(-\beta) R_x^{(2N-2)}(-g_U\phi_M-\gamma)
\nonumber\\
& ~~~ \times R_{zz}^{(2N-1,2N)}(g_U\phi_M) R_z^{(2N-1)}(g_U\phi_M)  R_x^{(2N)}(-\alpha) R_z^{(2N)}(-\beta) R_x^{(2N)}(-g_U\phi_M-\gamma) 
\nonumber\\
& ~~~ \times R_{zzz}^{(2N-2,2N-1,2N)}(\phi_M) R_x^{(2N-2)}(\gamma) R_z^{(2N-2)}(\beta) R_x^{(2N-2)}(\lambda+\alpha) R_x^{(2N)}(\gamma) R_z^{(2N)}(\beta) R_x^{(2N)}(\lambda+\alpha)
\nonumber\\
&~~~ \times R_{zzz}^{(2N-2,2N-1,2N)}(\phi_M)   R_x^{(2N-3)}(\alpha) R_x^{(2N-2)}(-\alpha) R_x^{(2N-1)}(\alpha) 
R_x^{(2N)}(-\alpha)
\nonumber\\
& ~~~ \times R_{zzz}^{(2N-3,2N-2,2N-1)}(\phi_M) R_x^{(2N-3)}(-\lambda-\alpha) R_z^{(2N-3)}(-\beta) R_x^{(2N-3)}(-\gamma) R_x^{(2N-1)}(-\lambda-\alpha) R_z^{(2N-1)}(-\beta) R_x^{(2N-1)}(-\gamma)
\nonumber\\
& ~~~ \times R_{zzz}^{(2N-3,2N-2,2N-1)}(\phi_M) R_x^{(2N-3)}(\gamma) R_z^{(2N-3)}(\beta) R_x^{(2N-3)}(\alpha) R_x^{(2N-1)}(\gamma) R_z^{(2N-1)}(\beta) R_x^{(2N-1)}(\alpha).
\label{Gate_Hubbard_W}
\end{align}

\begin{figure}[t]
\begin{center}
\includegraphics[width=1.0\textwidth]{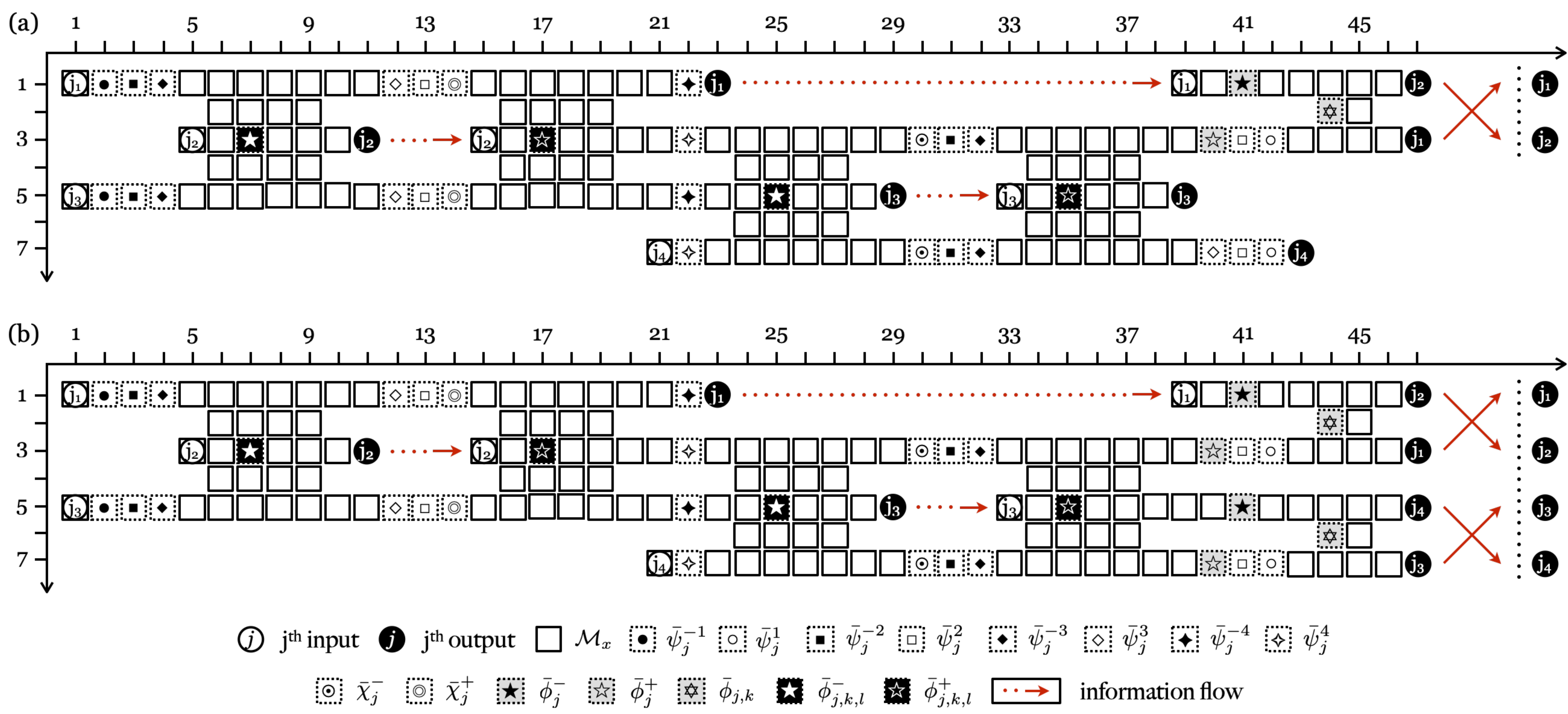}
\end{center}
\caption{Coordinate systems for assigning qubit positions in the measurement pattern for the Hubbard chain, which can be built by combining two types of measurement patterns for (a) $V_g^{(j)}$ ($1\leq j\leq N-2$) ($N\geq 3$), (b) $W_g^{(N-1)}$ ($N\geq 2$). Here, input and output qubits are indexed by (a) $(j_1, j_2, j_3, j_4) = (2j-1, 2j, 2j+1, 2j+2)$ for a given $j$, (b) $(j_1, j_2, j_3, j_4) = (2N-3, 2N-2, 2N-1, 2N)$ for a given $N$.
}
\label{fig_subroutine_hubbard}
\end{figure}

To implement the measurement pattern for $U_g$, we can take the same strategy as for the Kitaev chain. The procedure has two steps: First, we build two types of measurement patterns for $V_g^{(j)}$, $W_g^{(N-1)}$ [Eqs.~\eqref{Gate_Hubbard_V}, \eqref{Gate_Hubbard_W}] (Figure~\ref{fig_subroutine_hubbard}). Second, we combine them in a specific order for a given $N$. The resulting giant measurement pattern has a cascade structure flowing from the left top to the right bottom. Since detailed mathematical derivation is lengthy even for $N=2$, we just summarize the result below. 

Regarding the measurement pattern for $V_g^{(j)}$ [Figure~\ref{fig_subroutine_hubbard}(a)], it turns out that the measurement angles have the form: 
\begin{align}
& \bar{\phi}_{j,k,l}^{\pm} 
= -P_{\bar{\phi}_{j,k,l}^{\pm}}\phi_M,
~~
\bar{\phi}_{j,k} = -P_{\bar{\phi}_{j,k}}g_U\phi_M,
~~
\bar{\psi}_{j}^{r} 
= P_{\bar{\psi}_{j}^{r}}\psi^r,
\nonumber\\
& \bar{\phi}_{j}^{\pm} 
= \pm P_{\bar{\phi}_{j}^{\pm}}[g_U\phi_M + (1\pm 1)\gamma/2],
~~
\bar{\chi}_{j}^{\pm} 
= \pm P_{\bar{\chi}_{j}^{\pm}}(\lambda+\alpha),
\label{Measurement_angles_Hubbard}
\end{align}
where $\psi^r\in\{\pm\alpha,\pm\beta,\pm\gamma,\pm\lambda\}$ for $r=\pm1,\pm2,\pm3,\pm4$, and $P_\theta = (-1)^{S_\theta}$. The exponents for the three-qubit rotation gates are defined by
\begin{align}
S^{(j)}_{\bar{\phi}^{-}_{2j-1,2j,2j+1}} 
& = s_{(2,1)}^{(j)}+s_{(2,5)}^{(j)}+s_{(4,1)}^{(j)}+s_{(4,5)}^{(j)}+s_{(6,1)}^{(j)}+s_{(6,3)}^{(j)}+s_{(6,5)}^{(j)}+s_{(8,1)}^{(j)}+s_{(8,5)}^{(j)}+s_{(9,2)}^{(j)}+s_{(9,4)}^{(j)},
\\
S^{(j)}_{\bar{\phi}^{+}_{2j-1,2j,2j+1}} 
& = s_{(2,1)}^{(j)}+s_{(2,5)}^{(j)}+s_{(4,1)}^{(j)}+s_{(4,5)}^{(j)}+s_{(6,1)}^{(j)}+s_{(6,3)}^{(j)}+s_{(6,5)}^{(j)}+s_{(8,1)}^{(j)}+s_{(8,3)}^{(j)}+s_{(8,5)}^{(j)}+s_{(10,1)}^{(j)}+s_{(10,3)}^{(j)}+s_{(10,5)}^{(j)}
\nonumber\\
& ~ +s_{(12,1)}^{(j)}+s_{(12,5)}^{(j)}+s_{(14,1)}^{(j)}+s_{(14,5)}^{(j)}+s_{(16,1)}^{(j)}+s_{(16,3)}^{(j)}+s_{(16,5)}^{(j)}+s_{(18,1)}^{(j)}+s_{(18,5)}^{(j)}+s_{(19,2)}^{(j)}+s_{(19,4)}^{(j)},
\\
S^{(j)}_{\bar{\phi}^{-}_{2j,2j+1,2j+2}} 
& = s_{(2,5)}^{(j)}+s_{(4,5)}^{(j)}+s_{(6,3)}^{(j)}+s_{(6,5)}^{(j)}+s_{(7,2)}^{(j)}+s_{(8,1)}^{(j)}+s_{(8,5)}^{(j)}+s_{(9,4)}^{(j)}+s_{(10,1)}^{(j)}+s_{(10,3)}^{(j)}+s_{(12,1)}^{(j)}+s_{(14,1)}^{(j)}
\nonumber\\
& ~ +s_{(16,1)}^{(j)}+s_{(16,3)}^{(j)}+s_{(17,4)}^{(j)}+s_{(18,1)}^{(j)}+s_{(18,5)}^{(j)}+s_{(19,2)}^{(j)}+s_{(20,3)}^{(j)}+s_{(20,5)}^{(j)}+s_{(22,3)}^{(j)}+s_{(22,5)}^{(j)}+s_{(22,7)}^{(j)}
\nonumber\\
& ~ +s_{(24,3)}^{(j)}+s_{(24,5)}^{(j)}+s_{(24,7)}^{(j)}+s_{(26,3)}^{(j)}+s_{(26,7)}^{(j)}+s_{(27,4)}^{(j)}+s_{(27,6)}^{(j)},
\end{align}
\begin{align}
S^{(j)}_{\bar{\phi}^{+}_{2j,2j+1,2j+2}} 
& = s_{(2,5)}^{(j)}+s_{(4,5)}^{(j)}+s_{(6,3)}^{(j)}+s_{(6,5)}^{(j)}+s_{(7,2)}^{(j)}+s_{(8,1)}^{(j)}+s_{(8,5)}^{(j)}+s_{(9,4)}^{(j)}+s_{(10,1)}^{(j)}+s_{(10,3)}^{(j)}+s_{(12,1)}^{(j)}+s_{(14,1)}^{(j)}
\nonumber\\
& ~~~ +s_{(16,1)}^{(j)}+s_{(16,3)}^{(j)}+s_{(17,4)}^{(j)}+s_{(18,1)}^{(j)}+s_{(18,5)}^{(j)}+s_{(19,2)}^{(j)}+s_{(20,3)}^{(j)}+s_{(20,5)}^{(j)}+s_{(22,3)}^{(j)}+s_{(22,5)}^{(j)}+s_{(22,7)}^{(j)}
\nonumber\\
& ~~~ +s_{(24,3)}^{(j)}+s_{(24,5)}^{(j)}+s_{(24,7)}^{(j)}+s_{(26,3)}^{(j)}+s_{(26,5)}^{(j)}+s_{(26,7)}^{(j)}+s_{(28,3)}^{(j)}+s_{(28,5)}^{(j)}+s_{(28,7)}^{(j)}+s_{(30,3)}^{(j)}+s_{(30,7)}^{(j)}
\nonumber\\
& ~~~ +s_{(32,3)}^{(j)}+s_{(32,7)}^{(j)}+s_{(34,3)}^{(j)}+s_{(34,5)}^{(j)}+s_{(34,7)}^{(j)}+s_{(36,3)}^{(j)}+s_{(36,7)}^{(j)}+s_{(37,4)}^{(j)}+s_{(37,6)}^{(j)}.
\end{align}
The exponent for the two-qubit rotation gate is defined by
\begin{align}
S^{(j)}_{\bar{\phi}_{2j-1,2j}} 
& = s_{(2,1)}^{(j)}+s_{(4,1)}^{(j)}+s_{(6,1)}^{(j)}+s_{(6,3)}^{(j)}+s_{(7,4)}^{(j)}+s_{(8,1)}^{(j)}+s_{(8,5)}^{(j)}+s_{(9,2)}^{(j)}+s_{(10,3)}^{(j)}+s_{(10,5)}^{(j)}+s_{(12,5)}^{(j)}+s_{(14,5)}^{(j)}
\nonumber\\
& ~~~ +s_{(16,3)}^{(j)}+s_{(16,5)}^{(j)}+s_{(17,2)}^{(j)}+s_{(18,1)}^{(j)}+s_{(18,5)}^{(j)}+s_{(19,4)}^{(j)}+s_{(20,1)}^{(j)}+s_{(20,3)}^{(j)}+s_{(22,1)}^{(j)}+s_{(22,3)}^{(j)}+s_{(24,3)}^{(j)}
\nonumber\\
& ~~~ +s_{(25,4)}^{(j)}+s_{(26,5)}^{(j)}+s_{(27,6)}^{(j)}+s_{(28,7)}^{(j)}+s_{(30,7)}^{(j)}+s_{(32,7)}^{(j)}+s_{(34,7)}^{(j)}+s_{(35,6)}^{(j)}+s_{(36,5)}^{(j)}+s_{(37,4)}^{(j)}+s_{(38,3)}^{(j)}
\nonumber\\
& ~~~ +s_{(40,1)}^{(j)}+s_{(40,3)}^{(j)}+s_{(42,1)}^{(j)}+s_{(42,3)}^{(j)}+s_{(44,1)}^{(j)}+s_{(44,3)}^{(j)}+s_{(45,2)}^{(j)},
\end{align}
Other exponents for the single-qubit rotation gates are defined in the hierarchical form:
\begin{align}
& \left\{
\begin{array}{rl}
S^{(j)}_{\bar{\psi}^{-1}_{2j-1}} 
& = s_{(1,1)}^{(j)},
\\
S^{(j)}_{\bar{\psi}^{-3}_{2j-1}} 
& = S^{(j)}_{\bar{\psi}^{-1}_{2j-1}}+s_{(3,1)}^{(j)},
\\
S^{(j)}_{\bar{\psi}^3_{2j-1}} 
& = S^{(j)}_{\bar{\psi}^{-3}_{2j-1}}+s_{(5,1)}^{(j)}+s_{(6,2)}^{(j)}+s_{(7,3)}^{(j)}+s_{(8,4)}^{(j)}+s_{(9,5)}^{(j)}+s_{(11,5)}^{(j)},
\\
S^{(j)}_{\bar{\chi}^{+}_{2j-1}} 
& = S^{(j)}_{\bar{\psi}^3_{2j-1}}+s_{(13,5)}^{(j)},
\\
S^{(j)}_{\bar{\psi}^{-4}_{2j-1}} 
& = S^{(j)}_{\bar{\chi}^{+}_{2j-1}}+s_{(15,5)}^{(j)}+s_{(16,4)}^{(j)}+s_{(17,3)}^{(j)}+s_{(18,2)}^{(j)}+s_{(19,1)}^{(j)}+s_{(21,1)}^{(j)},
\end{array}
\right.
\label{S_hubbard_group_1}
\\
& \left\{
\begin{array}{rl}
S^{(j)}_{\bar{\psi}^{-2}_{2j-1}} 
& = s_{(2,1)}^{(j)},
\\
S^{(j)}_{\bar{\psi}^2_{2j-1}} 
& = S^{(j)}_{\bar{\psi}^{-2}_{2j-1}}+s_{(4,1)}^{(j)}+s_{(6,1)}^{(j)}+s_{(7,2)}^{(j)}+s_{(8,3)}^{(j)}+s_{(9,4)}^{(j)}+s_{(10,5)}^{(j)}+s_{(12,5)}^{(j)},
\\
S^{(j)}_{\bar{\phi}^{-}_{2j-1}} 
& = S^{(j)}_{\bar{\psi}^2_{2j-1}}+s_{(14,5)}^{(j)}+s_{(16,5)}^{(j)}+s_{(17,4)}^{(j)}+s_{(18,3)}^{(j)}+s_{(19,2)}^{(j)}+s_{(20,1)}^{(j)}+s_{(22,1)}^{(j)}+s_{(40,1)}^{(j)},
\end{array}
\right.
\\
& \left\{
\begin{array}{rl}
S^{(j)}_{\bar{\psi}^4_{2j}} 
& = s_{(5,3)}^{(j)}+s_{(6,2)}^{(j)}+s_{(6,4)}^{(j)}+s_{(7,1)}^{(j)}+s_{(7,3)}^{(j)}+s_{(7,5)}^{(j)}+s_{(8,2)}^{(j)}+s_{(8,4)}^{(j)}+s_{(9,3)}^{(j)}+s_{(15,3)}^{(j)}
\\
& ~~~ +s_{(16,2)}^{(j)}+s_{(16,4)}^{(j)}+s_{(17,1)}^{(j)}+s_{(17,3)}^{(j)}+s_{(17,5)}^{(j)}+s_{(18,2)}^{(j)}+s_{(18,4)}^{(j)}+s_{(19,3)}^{(j)}+s_{(21,3)}^{(j)},
\\
S^{(j)}_{\bar{\chi}^{-}_{2j}} 
& = S^{(j)}_{\bar{\psi}^4_{2j}}+s_{(23,3)}^{(j)}+s_{(24,4)}^{(j)}+s_{(25,5)}^{(j)}+s_{(26,6)}^{(j)}+s_{(27,7)}^{(j)}+s_{(29,7)}^{(j)},
\\
S^{(j)}_{\bar{\psi}^{-3}_{2j}} 
& = S^{(j)}_{\bar{\chi}^{-}_{2j}}+s_{(31,7)}^{(j)},
\\
S^{(j)}_{\bar{\phi}^{+}_{2j}} 
& = S^{(j)}_{\bar{\psi}^{-3}_{2j}}+s_{(33,7)}^{(j)}+s_{(34,6)}^{(j)}+s_{(35,5)}^{(j)}+s_{(36,4)}^{(j)}+s_{(37,3)}^{(j)}+s_{(39,3)}^{(j)},
\\
S^{(j)}_{\bar{\psi}^1_{2j}} 
& = S^{(j)}_{\bar{\phi}^{+}_{2j}}+s_{(41,3)}^{(j)},
\end{array}
\right.
\\
& \left\{
\begin{array}{rl}
S^{(j)}_{\bar{\psi}^{-2}_{2j}} 
& = s_{(6,3)}^{(j)}+s_{(7,2)}^{(j)}+s_{(7,4)}^{(j)}+s_{(8,1)}^{(j)}+s_{(8,3)}^{(j)}+s_{(8,5)}^{(j)}+s_{(9,2)}^{(j)}+s_{(9,4)}^{(j)}+s_{(10,3)}^{(j)}
\\
& ~~~ +s_{(16,3)}^{(j)}+s_{(17,2)}^{(j)}+s_{(17,4)}^{(j)}+s_{(18,1)}^{(j)}+s_{(18,3)}^{(j)}+s_{(18,5)}^{(j)}+s_{(19,2)}^{(j)}+s_{(19,4)}^{(j)}
\\
& ~~~ +s_{(20,3)}^{(j)}+s_{(22,3)}^{(j)}+s_{(24,3)}^{(j)}+s_{(25,4)}^{(j)}+s_{(26,5)}^{(j)}+s_{(27,6)}^{(j)}+s_{(28,7)}^{(j)}+s_{(30,7)}^{(j)},
\\
S^{(j)}_{\bar{\psi}^2_{2j}} 
& = S^{(j)}_{\bar{\psi}^{-2}_{2j}}+s_{(32,7)}^{(j)}+s_{(34,7)}^{(j)}+s_{(35,6)}^{(j)}+s_{(36,5)}^{(j)}+s_{(37,4)}^{(j)}+s_{(38,3)}^{(j)}+s_{(40,3)}^{(j)},
\end{array}
\right.
\\
& \left\{
\begin{array}{rl}
S^{(j)}_{\bar{\psi}^{-1}_{2j+1}} 
& = s_{(1,5)}^{(j)},
\\
S^{(j)}_{\bar{\psi}^{-3}_{2j+1}} 
& = S^{(j)}_{\bar{\psi}^{-1}_{2j+1}}+s_{(3,5)}^{(j)},
\\
S^{(j)}_{\bar{\psi}^3_{2j+1}} 
& = S^{(j)}_{\bar{\psi}^{-3}_{2j+1}}+s_{(5,5)}^{(j)}+s_{(6,4)}^{(j)}+s_{(7,3)}^{(j)}+s_{(8,2)}^{(j)}+s_{(9,1)}^{(j)}+s_{(11,1)}^{(j)},
\\
S^{(j)}_{\bar{\chi}^{+}_{2j+1}} 
& = S^{(j)}_{\bar{\psi}^3_{2j+1}}+s_{(13,1)}^{(j)},
\\
S^{(j)}_{\bar{\psi}^{-4}_{2j+1}} 
& = S^{(j)}_{\bar{\chi}^{+}_{2j+1}}+s_{(15,1)}^{(j)}+s_{(16,2)}^{(j)}+s_{(17,3)}^{(j)}+s_{(18,4)}^{(j)}+s_{(19,5)}^{(j)}+s_{(21,5)}^{(j)},
\end{array}
\right.
\\
& \left\{
\begin{array}{rl}
S^{(j)}_{\bar{\psi}^{-2}_{2j+1}} 
& = s_{(2,5)}^{(j)},
\\
S^{(j)}_{\bar{\psi}^2_{2j+1}} 
& = S^{(j)}_{\bar{\psi}^{-2}_{2j+1}}+s_{(4,5)}^{(j)}+s_{(6,5)}^{(j)}+s_{(7,4)}^{(j)}+s_{(8,3)}^{(j)}+s_{(9,2)}^{(j)}+s_{(10,1)}^{(j)}+s_{(12,1)}^{(j)},
\end{array}
\right.
\end{align}
\begin{align}
& \left\{
\begin{array}{rl}
S^{(j)}_{\bar{\psi}^4_{2j+2}} 
& = s_{(21,7)}^{(j)},
\\
S^{(j)}_{\bar{\chi}^{-}_{2j+2}} 
& = S^{(j)}_{\bar{\psi}^4_{2j+2}}+s_{(23,7)}^{(j)}+s_{(24,6)}^{(j)}+s_{(25,5)}^{(j)}+s_{(26,4)}^{(j)}+s_{(27,3)}^{(j)}+s_{(29,3)}^{(j)},
\\
S^{(j)}_{\bar{\psi}^{-3}_{2j+2}} 
& = S^{(j)}_{\bar{\chi}^{-}_{2j+2}}+s_{(31,3)}^{(j)},
\\
S^{(j)}_{\bar{\psi}^3_{2j+2}} 
& = S^{(j)}_{\bar{\psi}^{-3}_{2j+2}}+s_{(33,3)}^{(j)}+s_{(34,4)}^{(j)}+s_{(35,5)}^{(j)}+s_{(36,6)}^{(j)}+s_{(37,7)}^{(j)}+s_{(39,7)}^{(j)},
\\
S^{(j)}_{\bar{\psi}^1_{2j+2}} 
& = S^{(j)}_{\bar{\psi}^3_{2j+2}}+s_{(41,7)}^{(j)},
\end{array}
\right.
\\
& \left\{
\begin{array}{rl}
S^{(j)}_{\bar{\psi}^{-2}_{2j+2}} 
& = s_{(22,7)}^{(j)}+s_{(24,7)}^{(j)}+s_{(25,6)}^{(j)}+s_{(26,5)}^{(j)}+s_{(27,4)}^{(j)}+s_{(28,3)}^{(j)}+s_{(30,3)}^{(j)},
\\
S^{(j)}_{\bar{\psi}^2_{2j+2}} 
& = S^{(j)}_{\bar{\psi}^{-2}_{2j+2}}+s_{(32,3)}^{(j)}+s_{(34,3)}^{(j)}+s_{(35,4)}^{(j)}+s_{(36,5)}^{(j)}+s_{(37,6)}^{(j)}+s_{(38,7)}^{(j)}+s_{(40,7)}^{(j)}.
\end{array}
\right.
\label{S_hubbard_group_8}
\end{align}
We also find the total byproduct operator:
\begin{align}
U_\Sigma^{(j)} 
& = Z_{2j-1}^{S^{(j)}_{Z_{2j-1}}} X_{2j-1}^{S^{(j)}_{X_{2j-1}}} Z_{2j}^{S^{(j)}_{Z_{2j}}} X_{2j}^{S^{(j)}_{X_{2j}}} Z_{2j+1}^{S^{(j)}_{Z_{2j+1}}} X_{2j+1}^{S^{(j)}_{X_{2j+1}}} Z_{2j+2}^{S^{(j)}_{Z_{2j+2}}} X_{2j+2}^{S^{(j)}_{X_{2j+2}}},
\end{align}
where the exponents are defined by using Eqs.~\eqref{S_hubbard_group_1}-\eqref{S_hubbard_group_8}:
\begin{align}
S^{(j)}_{Z_{2j-1}} 
& = S^{(j)}_{\bar{\psi}^{-4}_{2j-1}}+s_{(39,1)}^{(j)}+s_{(41,1)}^{(j)}+s_{(43,1)}^{(j)}+s_{(44,2)}^{(j)}+s_{(45,3)}^{(j)},
\\
S^{(j)}_{X_{2j-1}} 
& = S^{(j)}_{\bar{\phi}^{-}_{2j-1}}+s_{(42,1)}^{(j)}+s_{(44,1)}^{(j)}+s_{(45,2)}^{(j)}+s_{(46,3)}^{(j)},
\\
S^{(j)}_{Z_{2j}} 
& = S^{(j)}_{\bar{\psi}^1_{2j}}+s_{(43,3)}^{(j)}+s_{(44,2)}^{(j)}+s_{(45,1)}^{(j)},
\\
S^{(j)}_{X_{2j}} 
& = S^{(j)}_{\bar{\psi}^2_{2j}}+s_{(42,3)}^{(j)}+s_{(44,3)}^{(j)}+s_{(45,2)}^{(j)}+s_{(46,1)}^{(j)},
\\
S^{(j)}_{Z_{2j+1}} 
& = S^{(j)}_{\bar{\psi}^{-4}_{2j+1}}+s_{(23,5)}^{(j)}+s_{(24,4)}^{(j)}+s_{(24,6)}^{(j)}+s_{(25,3)}^{(j)}+s_{(25,5)}^{(j)}+s_{(25,7)}^{(j)}+s_{(26,4)}^{(j)}+s_{(26,6)}^{(j)}+s_{(27,5)}^{(j)}+s_{(33,5)}^{(j)}+s_{(34,4)}^{(j)}
\nonumber\\
& ~~~ +s_{(34,6)}^{(j)}+s_{(35,3)}^{(j)}+s_{(35,5)}^{(j)}+s_{(35,7)}^{(j)}+s_{(36,4)}^{(j)}+s_{(36,6)}^{(j)}+s_{(37,5)}^{(j)},
\label{S_Z_2j+1}
\\
S^{(j)}_{X_{2j+1}} 
& = S^{(j)}_{\bar{\psi}^2_{2j+1}}+s_{(14,1)}^{(j)}+s_{(16,1)}^{(j)}+s_{(17,2)}^{(j)}+s_{(18,3)}^{(j)}+s_{(19,4)}^{(j)}+s_{(20,5)}^{(j)}+s_{(22,5)}^{(j)}+s_{(24,5)}^{(j)}+s_{(25,4)}^{(j)}+s_{(25,6)}^{(j)}+s_{(26,3)}^{(j)}
\nonumber\\
& ~~~ +s_{(26,5)}^{(j)}+s_{(26,7)}^{(j)}+s_{(27,4)}^{(j)}+s_{(27,6)}^{(j)}+s_{(28,5)}^{(j)}+s_{(34,5)}^{(j)}+s_{(35,4)}^{(j)}+s_{(35,6)}^{(j)}+s_{(36,3)}^{(j)}+s_{(36,5)}^{(j)}+s_{(36,7)}^{(j)}+s_{(37,4)}^{(j)}
\nonumber\\
& ~~~ +s_{(37,6)}^{(j)}+s_{(38,5)}^{(j)},
\label{S_X_2j+1}
\\
S^{(j)}_{Z_{2j+2}} 
& = S^{(j)}_{\bar{\psi}^1_{2j+2}},
\label{S_Z_2j+2}
\\
S^{(j)}_{X_{2j+2}} 
& = S^{(j)}_{\bar{\psi}^2_{2j+2}}+s_{(42,7)}^{(j)}.
\label{S_X_2j+2}
\end{align}
The same strategy is applied to $W_g^{(N-1)}$ [Figure~\ref{fig_subroutine_hubbard}(b)]. Due to structural similarity, most of the results for $V_g^{(j)}$ can be held only with the index $j$ replaced by $N-1$. The exceptions occur for Eqs.~\eqref{S_Z_2j+1}-\eqref{S_X_2j+2} with modifications:
\begin{align}
\tilde{S}^{(N-1)}_{Z_{2N-1}} 
& = S^{(N-1)}_{Z_{2N-1}} + s_{(39,5)}^{(N-1)} + s_{(41,5)}^{(N-1)} + s_{(43,5)}^{(N-1)} + s_{(44,6)}^{(N-1)} + s_{(45,7)}^{(N-1)},
\\
\tilde{S}^{(N-1)}_{X_{2N-1}} 
& = S^{(N-1)}_{X_{2N-1}} + s_{(40,5)}^{(N-1)} + s_{(42,5)}^{(N-1)} + s_{(44,5)}^{(N-1)} + s_{(45,6)}^{(N-1)} + s_{(46,7)}^{(N-1)},
\\
\tilde{S}^{(N-1)}_{Z_{2N}} 
& = S^{(N-1)}_{Z_{2N}} + s_{(43,7)}^{(N-1)} + s_{(44,6)}^{(N-1)} + s_{(45,5)}^{(N-1)},
\\
\tilde{S}^{(N-1)}_{X_{2N}} 
& = S^{(N-1)}_{X_{2N}} + s_{(44,7)}^{(N-1)} + s_{(45,6)}^{(N-1)} + s_{(46,5)}^{(N-1)}.
\end{align}
In the right bottom of Figure~\ref{fig_subroutine_hubbard}(b), three measurement angles $(\bar{\phi}^{-}_{2N-1}, \bar{\phi}^{+}_{2N}, \bar{\phi}_{2N-1,2N})$ are introduced with definitions:
\begin{align}
\bar{\phi}_{2N-1,2N} 
= -P_{\bar{\phi}_{2N-1,2N}} g_U\phi_M,
~~
\bar{\phi}^{-}_{2N-1} 
= -P_{\bar{\phi}^{-}_{2N-1}} g_U\phi_M,
~~
\bar{\phi}^{+}_{2N} 
= P_{\bar{\phi}^{+}_{2N}}(g_U\phi_M + \gamma),
\end{align}
where the exponents are defined by
\begin{align}
S^{(N-1)}_{\bar{\phi}^{-}_{2N-1}} 
& = S^{(N-1)}_{X_{2N-1}}+s_{(40,5)}^{(N-1)},
\\
S^{(N-1)}_{\bar{\phi}^{+}_{2N}} 
& = S^{(N-1)}_{\bar{\psi}^{-3}_{2N-2}}+s_{(33,3)}^{(N-1)}+s_{(34,4)}^{(N-1)}+s_{(35,5)}^{(N-1)}+s_{(36,6)}^{(N-1)}+s_{(37,7)}^{(N-1)}+s_{(39,7)}^{(N-1)},
\end{align}
\begin{align}
S^{(N-1)}_{\bar{\phi}_{2N-1,2N}} 
& = s_{(2,5)}^{(N-1)}+s_{(4,5)}^{(N-1)}+s_{(6,5)}^{(N-1)}+s_{(7,4)}^{(N-1)}+s_{(8,3)}^{(N-1)}+s_{(9,2)}^{(N-1)}+s_{(10,1)}^{(N-1)}+s_{(12,1)}^{(N-1)}+s_{(14,1)}^{(N-1)}+s_{(16,1)}^{(N-1)}+s_{(17,2)}^{(N-1)}
\nonumber\\
& ~~~ +s_{(18,3)}^{(N-1)}+s_{(19,4)}^{(N-1)}+s_{(20,5)}^{(N-1)}+s_{(22,5)}^{(N-1)}+s_{(22,7)}^{(N-1)}+s_{(24,5)}^{(N-1)}+s_{(24,7)}^{(N-1)}+s_{(25,4)}^{(N-1)}+s_{(26,3)}^{(N-1)}+s_{(26,7)}^{(N-1)}+s_{(27,6)}^{(N-1)}
\nonumber\\
& ~~~ +s_{(28,5)}^{(N-1)}+s_{(28,3)}^{(N-1)}+s_{(30,3)}^{(N-1)}+s_{(32,3)}^{(N-1)}+s_{(34,3)}^{(N-1)}+s_{(34,5)}^{(N-1)}+s_{(35,6)}^{(N-1)}+s_{(36,3)}^{(N-1)}+s_{(36,7)}^{(N-1)}+s_{(37,4)}^{(N-1)}+s_{(38,5)}^{(N-1)}
\nonumber\\
& ~~~ +s_{(38,7)}^{(N-1)}+s_{(40,5)}^{(N-1)}+s_{(40,7)}^{(N-1)} +s_{(42,5)}^{(N-1)}+s_{(42,7)}^{(N-1)}+s_{(44,5)}^{(N-1)}+s_{(44,7)}^{(N-1)}+s_{(45,6)}^{(N-1)}.
\end{align}

\begin{figure}[t]
\begin{center}
\includegraphics[width=1.0\textwidth]{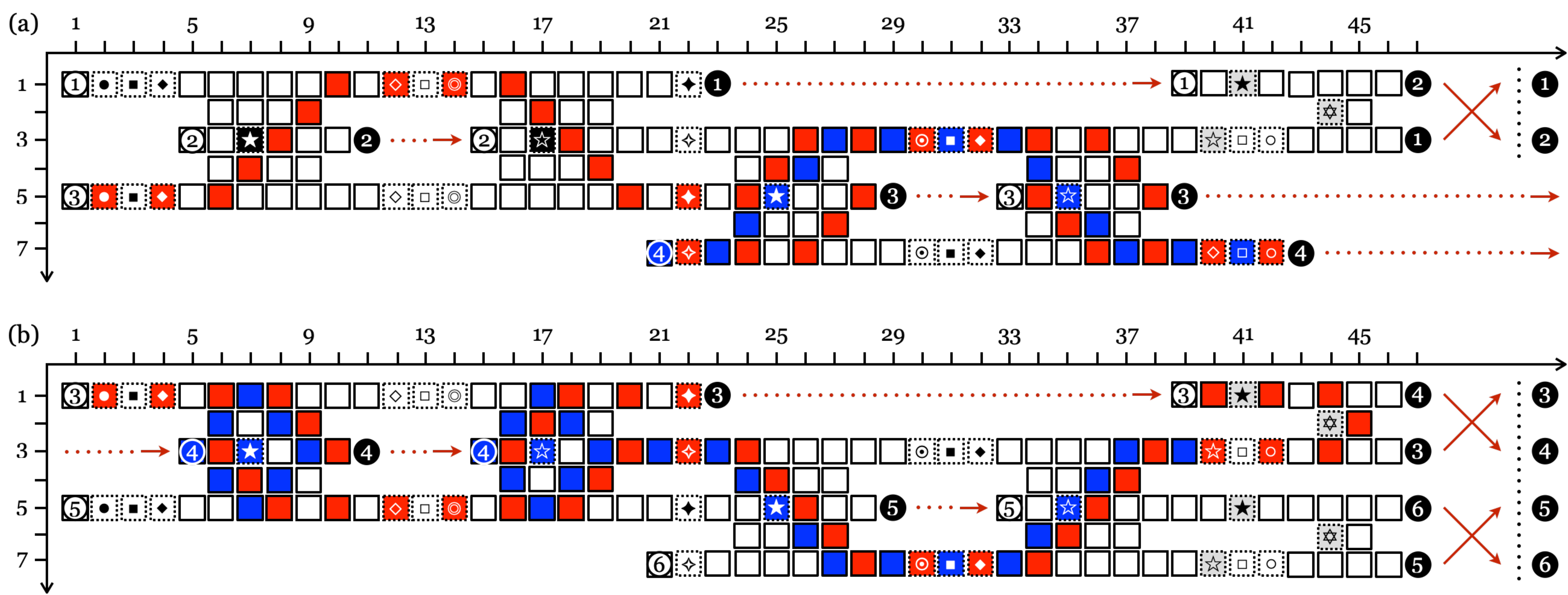}
\end{center}
\caption{Schematic for correlation centers contributing to $S^{(2)}_{\bar{\phi}^{+}_{4}}$ (blue boxes) and $S^{(2)}_{\bar{\phi}_{3,4}}$ (red boxes) in the giant measurement pattern for the $N=3$ Hubbard chain. Here, correlation centers are highlighted over two successive measurement patterns for (a) $V_g^{(1)}$ and (b) $W_g^{(2)}$.}
\label{fig_correlation_hubbard}
\end{figure}

To complete the concatenation process for $N\geq 3$, we need to reapply Pauli propagation to push all byproduct operators to the left side of all other unitary gates in the giant measurement pattern. In this process, the exponents in $V_g^{(j)}$ are adjusted to accumulate further the measurement outcomes in $V_g^{(j-1)}$. The same thing also happens for the pair: $\{W_g^{(N-1)}, V_g^{(N-2)}\}$. Consequently, some exponents are modified in the following way:
\begin{align}
S^{(j)}_{\bar{\phi}^{-}_{2j-1,2j,2j+1}} 
& \rightarrow S^{(j)}_{\bar{\phi}^{-}_{2j-1,2j,2j+1}} + S^{(j-1)}_{X_{2j-1}} + S^{(j-1)}_{X_{2j}},
\\
S^{(j)}_{\bar{\phi}^{+}_{2j-1,2j,2j+1}} 
& \rightarrow S^{(j)}_{\bar{\phi}^{+}_{2j-1,2j,2j+1}} + S^{(j-1)}_{X_{2j-1}} + S^{(j-1)}_{X_{2j}},
\\
S^{(j)}_{\bar{\phi}_{2j-1,2j}} 
& \rightarrow S^{(j)}_{\bar{\phi}_{2j-1,2j}} + S^{(j-1)}_{X_{2j-1}} + S^{(j-1)}_{X_{2j}},
\\
S^{(j)}_{\theta_{2j-1}} 
& \rightarrow S^{(j)}_{\theta_{2j-1}} + S^{(j-1)}_{Z_{2j-1}},
\\
S^{(j)}_{\theta_{2j}} 
& \rightarrow S^{(j)}_{\theta_{2j}} + S^{(j-1)}_{Z_{2j}},
\\
S^{(j)}_{\vartheta_{2j-1}} 
& \rightarrow S^{(j)}_{\vartheta_{2j-1}} + S^{(j-1)}_{X_{2j-1}},
\\
S^{(j)}_{\vartheta_{2j}} 
& \rightarrow S^{(j)}_{\vartheta_{2j}} + S^{(j-1)}_{X_{2j}},
\end{align}
for $\theta_j\in\{\bar{\psi}^{\pm1}_j, \bar{\psi}^{\pm3}_j, \bar{\psi}^{\pm4}_j,  \bar{\chi}^{\pm}_j, \bar{\phi}^{+}_j\}$ and $\vartheta_j\in\{\bar{\psi}^{\pm2}_j, \bar{\phi}^{-}_j\}$. Figure~\ref{fig_correlation_hubbard} demonstrates the positions of correlation centers contributing to $S^{(2)}_{\bar{\phi}^{+}_{4}}$ and $S^{(2)}_{\bar{\phi}_{3,4}}$ in the giant measurement pattern for the $N=3$ Hubbard chain.

This concludes our derivation of expressions discussed in the main text:  (1) the signs for the measurement angles, $S_\theta^\textrm{H}$, and (2) the byproduct operators for the Hubbard chain.  We have also proven the statement that the measurement pattern for the Hubbard chain shown in the main text can be concatenated for larger chains.

\newpage
\section{Estimation of Trotter steps and measurement precision}

\begin{figure}[t]
\begin{center}
\includegraphics[width=0.85\textwidth]{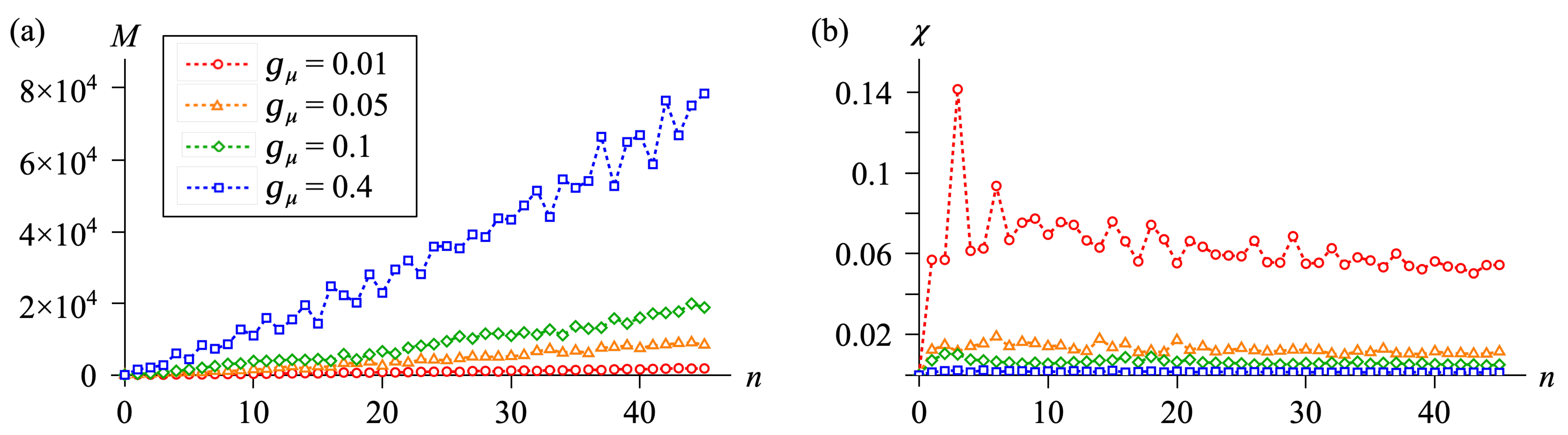}
\end{center}
\caption{
(a) Minimum number of Trotter steps, $M$, for time step $n(=0,1,\cdots,L-1)$. Here, we consider the Kitaev chain, and set $\eta/w = 0.02$, $\delta\omega/w = 0.01$, $L=46$, $N=4$, and $g_{\mu}=0.01$ (red), 0.05 (orange), 0.1 (green), 0.4 (blue). Convergence of solutions is achieved within a tolerance $\delta_\textrm{T}=10^{-2}$.
(b) Normalized measurement angle $\chi_n$ corresponding to the minimum $M$ found for each $n$.}
\label{fig_measurement_precision}
\end{figure}

In this section, we explicitly show the empirical calculation used to obtain the minimum number of Trotter steps, $M$, and the normalized measurement angle $\chi_n\equiv nw/(\delta\omega L M)$ discussed in the main text.  Figure~\ref{fig_measurement_precision}(a) plots the minimum value of $M$ needed to meet tolerances for the Kitaev chain for several different values of the chemical potential.  We see that the $M$ needed increases nearly linearly with the time step index.  Figure~\ref{fig_measurement_precision}(b) plots the corresponding measurement angles needed as a function of time step for several different chemical potentials.  Here we see that, as stated in the main text, the largest measurement angle is below $2\pi$ (as needed), and that the smallest angle needed to be measured can become very small as $g_{\mu} \rightarrow 1$.  These graphs show how the bounds on $M$ and $\chi_n$ stated in the main text were obtained.

\section{Estimation of resource requirements}

In this section, we prove how we obtained the resources requirements for a single time step of Eq.~(8) in the main text and for fixed $M$ and $N$, shown in Table I in the main text. 

\subsection{Kitaev chain}

$\bullet$ SLCS measurements: On an SLCS for a Kitaev chain (Figs.~\ref{fig_subroutine_kitaev_N2} and \ref{fig_subroutine_kitaev_N3}), all Pauli-$x$ (open boxes) and adaptive measurements (boxes including symbols) are counted. For our purpose, we don't count measurements on input qubits lying at the end of information flow. For $N=2$, we consider Fig.~\ref{fig_subroutine_kitaev_N2}(a). The measurement count is $24$. For $N=3$, Figs.~\ref{fig_subroutine_kitaev_N3}(a) and (c) with $(N_1,N_2) = (2,3)$ are combined. The measurement counts are $20$ and $21$, respectively. The total measurement count is $41$. For $N\geq 4$, the $N-3$ copies of Fig.~\ref{fig_subroutine_kitaev_N3}(b), indexed by $j$ ($2\leq j\leq N-2$), respectively, are stacked side by side, and then surrounded by Figs.~\ref{fig_subroutine_kitaev_N3}(a) and (c) with $(N_1,N_2) = (N-1,N)$. The measurement count for Fig.~\ref{fig_subroutine_kitaev_N3}(b) is $17$. The total measurement count is then  $20 + 17(N-3) + 21 = 17N - 10$. $M$-times repetition produces $(17N - 10)M$.

$\bullet$ CCS measurements: On a CCS, all Pauli-$x$ measurements in the body section of SLCS are excluded. As before, we don't count measurements on input qubits lying at the end of information flow. For $N=2$, the measurement count in Fig.~\ref{fig_subroutine_kitaev_N2}(a) is reduced to $13$. For $N=3$, the measurement counts in Figs.~\ref{fig_subroutine_kitaev_N3}(a) and (c) are reduced to $10$ and $11$, respectively. Merging Figs.~\ref{fig_subroutine_kitaev_N3}(a) and (c), and further excluding Pauli-$x$ measurement at input qubit $2$ in Fig.~\ref{fig_subroutine_kitaev_N3}(c), the total measurement count is reduced to $10 + 11 - 1 = 20$. For $N\geq 4$, the measurement count for Fig.~\ref{fig_subroutine_kitaev_N3}(b) is reduced to $8$. Merging Figs.~\ref{fig_subroutine_kitaev_N3}(a)-(c) in the same way as before, and further excluding Pauli-$x$ measurements at input qubits $j$ ($2\leq j\leq N-1$) in Figs.~\ref{fig_subroutine_kitaev_N3}(b) and (c), the total measurement count is reduced to $10 + 8(N-3) + 11 - (N-2) = 7N-1$. $M$-times repetition produces $(7N-1)M$.


$\bullet$ Circuit-based gates: Circuit-based gates are counted by using Eqs.~\eqref{Gate_Kitaev_N2}-\eqref{Gate_Kitaev_WN}. It turns out that the result is consistent with the CCS measurement count.


\subsection{Hubbard chain}

$\bullet$ SLCS measurements: On an SLCS for a Hubbard chain (Fig.~\ref{fig_subroutine_hubbard}), all Pauli-$x$ (open boxes) and adaptive measurements (boxes including symbols) are counted. For our purpose, we don't count measurements on input qubits lying at the end of information flow. For $N=2$, we consider only Fig.~\ref{fig_subroutine_hubbard}(b). The  measurement count is $168$. For $N\geq 3$, the $N-2$ copies of Fig.~\ref{fig_subroutine_hubbard}(a), indexed by $j$ ($1\leq j\leq N-2$), respectively, are stacked side by side, and we end up with Fig.~\ref{fig_subroutine_hubbard}(b) at the rightmost side. The measurement count for Fig.~\ref{fig_subroutine_hubbard}(a) is $154$ for $j=1$ or $156$ for $2\leq j\leq N-2$. The measurement count for Fig.~\ref{fig_subroutine_hubbard}(b) is $170$. The total measurement count is $154 + 156(N-3) + 170 = 156N-144$. $M$-times repetition produces $(156N-144)M$.


$\bullet$ CCS measurements: On a CCS, all Pauli-$x$ measurements in the body section of SLCS are excluded. As before, we don't count measurements on input qubits lying at the end of information flow. For $N=2$, the measurement count in Fig.~\ref{fig_subroutine_hubbard}(b) is reduced to $36$. For $N\geq 3$, the measurement count for Fig.~\ref{fig_subroutine_hubbard}(a) is reduced to $34$ for $j=1$ or $36$ for $2\leq j\leq N-2$. The measurement count for Fig.~\ref{fig_subroutine_hubbard}(b) is reduced to $38$. Merging Figs.~\ref{fig_subroutine_hubbard}(a) and (b) in the same way as before, and further excluding Pauli-$x$ measurements on input qubits not lying at the end of information flow, the total measurement count is reduced to $34 + 36(N-3) + 38 - 2(N-2) = 34N - 32$. $M$-times repetition produces $(34N-32)M$.


$\bullet$ Circuit-based gates: Circuit-based gates are counted by using Eqs.~\eqref{Gate_Hubbard_N}-\eqref{Gate_Hubbard_W}. It turns out that the result is consistent with the CCS measurement count.


\section{Tolerable errors in measurement angles}




In this section, we prove that, in a hybrid quantum eigenvalue estimation algorithm, peak centers are intact while peak weights are shifted for certain types of measurement errors.  For the purpose of demonstration, we revisit the measurement pattern for the $N=2$ Kitaev chain, and consider error $\varepsilon_\theta$ in Euler-decomposed measurement angle $\theta$ due to inaccurate measurements:
\begin{align}
\bar{\psi}^{-1}_{j} = -P_{\bar{\psi}^{-1}_{j})} (\alpha + \varepsilon_{\bar{\psi}^{-1}_{j}}),
& ~~~
\bar{\psi}^{-2}_{j} = -P_{\bar{\psi}^{-2}_{j}} (\beta + \varepsilon_{\bar{\psi}^{-2}_{j}}),
~~~
\bar{\phi}_{j} = -P_{\bar{\phi}_j}(2g_\mu\phi_M + \gamma + \varepsilon_{\bar{\phi}_{j}}),
\label{Measurement_angle_Kitaev_error_filled}
\\
\bar{\psi}^{1}_{j} = P_{\bar{\psi}^{1}_{j}} (\alpha + \varepsilon_{\bar{\psi}^{1}_{j}}),
& ~~~
\bar{\psi}^{2}_{j} = P_{\bar{\psi}^{2}_{j}} (\beta + \varepsilon_{\bar{\psi}^{2}_{j}}),
~~~
\bar{\psi}^{3}_{j} = P_{\bar{\psi}^{3}_{j}} (\gamma + \varepsilon_{\bar{\psi}^{3}_{j}}),
\label{Measurement_angle_Kitaev_error_unfilled}
\end{align}
where $-\alpha = \beta = \gamma = \pi/2$ and $j\in\{1,2\}$. For clarity, we assume that no error is invoked by Pauli-$x$ measurements (preserving stabilizer) and quantum state tomography of input and output qubits. Plugging Eqs.~\eqref{Measurement_angle_Kitaev_error_filled} and \eqref{Measurement_angle_Kitaev_error_unfilled} in Eq.~\eqref{Gate_Kitaev_combined_N_2}, and applying Pauli propagation to unitary gates and byproduct operators, the error-prone output wavefunction $|\psi_\textrm{O}^\varepsilon\rangle$ has the following form in the first-order Trotter-Suzuki decomposition:
\begin{align}
|\psi_\textrm{O}^\varepsilon\rangle = (\tilde{U}_\varepsilon^\dag U_g U_\varepsilon)^M |\psi_\textrm{I}\rangle,
\label{output_kitaev_N2_error}
\end{align}
up to the byproduct operator. Here, $U_g$ is defined by Eq.~\eqref{Gate_Kitaev_N2}, and the error-prone parts of unitary gates are defined by
\begin{align}
U_\varepsilon 
& = R_z^{(1)}(-\varepsilon_{\bar{\phi}_{1}}) R_y^{(1)}(-\varepsilon_{\bar{\psi}^{-2}_{1}}) R_x^{(1)}(\varepsilon_{\bar{\psi}^{-1}_{1}})
R_z^{(2)}(-\varepsilon_{\bar{\phi}_{2}}) R_y^{(2)}(-\varepsilon_{\bar{\psi}^{-2}_{2}}) R_x^{(2)}(\varepsilon_{\bar{\psi}^{-1}_{2}}),
\\
\tilde{U}_\varepsilon
& = R_z^{(1)}(-\varepsilon_{\bar{\psi}^{3}_{1}}) R_y^{(1)}(-\varepsilon_{\bar{\psi}^{2}_{1}})
R_x^{(1)}(\varepsilon_{\bar{\psi}^{1}_{1}}) R_z^{(2)}(-\varepsilon_{\bar{\psi}^{3}_{2}})
R_y^{(2)}(-\varepsilon_{\bar{\psi}^{2}_{2}}) R_x^{(2)}(\varepsilon_{\bar{\psi}^{1}_{2}}).
\end{align}

We now discuss the impact of errors on spectral properties. Two different scenarios are available depending on the condition: (i) $U_\varepsilon = \tilde{U}_\varepsilon$, (ii) $U_\varepsilon \neq \tilde{U}_\varepsilon$. First, we notice that the symmetric condition (i) requires a special measurement protocol: errors in any of the first three filled symbols in Figure~\ref{fig_subroutine_kitaev_N2}(a) should match the errors in the next three unfilled symbols in the reverse order. Below, we show that errors are effectively mitigated in this case. We start by making an overlap between the input and output wavefunctions:
\begin{align}
\langle \psi_\textrm{I} |\psi_\textrm{O}^\varepsilon\rangle
= \langle \psi_\textrm{I} | (U_\varepsilon^\dag U_g U_\varepsilon)^M |\psi_\textrm{I}\rangle 
= \langle \psi_\textrm{I}^\varepsilon | e^{-iH_\textrm{K}t} |\psi_\textrm{I}^\varepsilon\rangle.
\label{Overlap_function_Kitaev_error}
\end{align}
Noticeably, in Eq.~\eqref{Overlap_function_Kitaev_error}, errors in measurement angles were effectively shifted to perturbations to input qubits: $|\psi_\textrm{I}^\varepsilon\rangle \equiv U_\varepsilon^M |\psi_\textrm{I}\rangle$, while the original time evolution operator was recovered: $e^{-iH_\textrm{K}t} = U_g^M$. It is understood that such perturbations are tolerable, because $\vert \psi_{\text{I}}\rangle$ only needs a non-zero overlap with the exact ground state. Expanding the input wavefunction into $|\psi_\textrm{I}\rangle = \sum_j c_j |u_j\rangle$ with the eigenstate $|u_j\rangle$ satisfying $H_\textrm{K} |u_j\rangle = \mathcal{E}_j |u_j\rangle$, Eq.~\eqref{Overlap_function_Kitaev_error} is written in the spectral representation:
\begin{align}
\langle\psi_\textrm{I} |\psi_\textrm{O}^\varepsilon\rangle = \sum_j |d_j^\varepsilon|^2 e^{-i\mathcal{E}_j t},
\label{Overlap_function_Kitaev_error_spectral}
\end{align}
where we define the modified expansion coefficients:
\begin{align}
d_j^\varepsilon = \sum_{j'} \langle u_j| U_\varepsilon^M |u_{j'}\rangle c_{j'},
\end{align}
which are the mixture of the original coefficients $c_j$, mediated by $U_\varepsilon$. Plugging Eq.~\eqref{Overlap_function_Kitaev_error_spectral} in Eq.~(2) (in the main text) gives rise to the formula:
\begin{align}
\mathcal{A}_\varepsilon(\omega) = \sum_j |d_j^\varepsilon|^2 A_j(\omega),
\label{spectral_function_Kitaev_error}
\end{align}
where the Lorentzian peaks are defined by $A_j(\omega) = -(1/\pi) {\rm Im} [1/(\omega - \mathcal{E}_j + i\eta)]$. The structure of Eq.~\eqref{spectral_function_Kitaev_error} shows that errors in measurement angles are tolerable in the symmetric condition: energy eigenvalues are intact while peak weights are shifted.

In the asymmetric condition (ii), there is no way to reconstruct $e^{-iH_\textrm{K}t}$ from Eq.~\eqref{output_kitaev_N2_error}, because $\tilde{U}_\varepsilon^\dag U_\varepsilon \neq \mathbb{I}$, $[U_g,U_\varepsilon]\neq 0$, and $[U_g,\tilde{U}_\varepsilon^\dag]\neq 0$. Consequently, energy eigenvalues are not intact (pole structures are not preserved), that is, general errors are not tolerable.

Finally, we note that error in measurement angle $\phi_M$ is also tolerable. It turns out that such error invokes global shifts in the time interval, and does not impact the Fourier transform.